\documentclass[acmsmall,screen]{acmart}
\setcopyright{none}
\copyrightyear{2026}
\acmYear{2026}
\acmISBN{}
\acmDOI{}

\usepackage{fancyhdr}

\usepackage{amsmath,array,subcaption,xspace,algpseudocodex,minted,longtable,multirow}
\usepackage{pifont}

\def\emptyset{\varnothing}
\def\phi{\varphi}
\def\psi{\varpsi}

\def\N{\mathbb{N}}
\def\V{\mathbb{V}}
\def\C{\mathbb{C}}
\def\E{\mathcal{E}}

\def\Rocq{\textsc{Rocq}\xspace}
\def\OCaml{\textsc{OCaml}\xspace}
\def\OCanren{\textsc{OCanren}\xspace}
\def\miniKanren{\textsc{miniKanren}\xspace}

\newcommand{\textfn}[1]{\textsf{\slshape #1}}
\def\arity{\textfn{arity}\xspace}
\def\subterms{\textfn{subterms}\xspace}
\def\dom{\textfn{dom}\xspace}
\def\lhs{\textfn{lhs}\xspace}
\def\repr{\textfn{repr}\xspace}
\def\rhs{\textfn{rhs}\xspace}
\def\walk{\textfn{walk}\xspace}
\def\union{\textfn{union}\xspace}
\def\commonpart{\textfn{common\_part}\xspace}
\def\unify{\textfn{unify}\xspace}

\title{Efficient Rational Unification for \miniKanren}

\author{Eridan Domoratskiy}
\email{eridan200@mail.ru}
\orcid{0009-0005-9020-3403}

\author{Dmitry Boulytchev}
\email{dboulytchev@math.spbu.ru}
\orcid{0000-0001-8363-7143}

\affiliation{%
  \institution{St.~Petersburg State University}
  \city{St.~Petersburg}  
  \country{Russia}
}



\begin{document}

\settopmatter{printacmref=false}
\settopmatter{printfolios=true}
\renewcommand\footnotetextcopyrightpermission[1]{}
\pagestyle{fancy}
\fancyfoot{}
\fancyfoot[R]{miniKanren'26}
\fancypagestyle{firstfancy}{
  \fancyhead{}
  \fancyhead[R]{miniKanren'26}
  \fancyfoot{}
}
\makeatletter
\let\@authorsaddresses\@empty
\makeatother

\begin{abstract}
  We present an efficient algorithm for rational term unification in persistent settings
  which demonstrates a comparable performance w.r.t. the conventional \miniKanren
  unification with triangular substitution for Herbrand terms. Our algorithm is based
  on existing Martelli-Rossi approach and uses some adjustments to make the implementation
  more conventional. We provide certified proofs of principal algorithm properties
  in the \Rocq proof assistant and showcase the results of a comprehensive performance
  evaluation.
\end{abstract}

\maketitle
\thispagestyle{firstfancy}
\section{Introduction}

In \miniKanren, we conventionally deal with only finite terms (Herbrand trees) such as lists or trees but sometimes it is convenient to consider recursively defined terms such as streams or recursive types, named rational terms. In practical logic programming, algorithms for first-order rational unification are also used to reduce the time consumption by omitting the ``occurs check''.

There are some useful applications for \miniKanren with rational unification support: type inference with recursive types~\cite{domoratskiy2024relational}, path finding in cyclic graphs, etc~\cite{simon2006coinductive}. For instance, one may implement an STLC type checker in \miniKanren and achieve a solver for type checking, type inference and type inhabitation problems. Being run under a \miniKanren with rational unification, these solvers implicitly extend to support either recursive types, recursive terms (where recursion may be treated as an application of the \textbf{Y}-combinator under the call-by-name evaluation strategy) or both of them simultaneously (depending on a query).

However, this support stays partial since in some cases such a solver may try to descend recursively into a rational term which will cause it to go into an infinite loop. These problems may be addressed using special descending guards such as ``inductive''~\cite{bol1991analysis} and ``coinductive''~\cite{simon2006coinductive} relation annotations, and their concrete usage scenarios is a topic for future research. Also, these guards may allow the user to adjust the terms' domain depending on the concrete problem~\cite{dagnino2020flexible}.

Despite the existence of a variety of rational unification algorithms, in practice only the Huet's one~\cite{huet1976resolution} is used.
While it shows brilliant results in many applications, we show that it is inappropriate to use in persistent rational unification which
is required to implement \miniKanren due to the interleaving search.

We present our own algorithm based on the Martelli and Rossi's work~\cite{martelli1984efficient} which uses an interesting approach to avoid infinite looping by reducing the size of intermediate state.
Our work includes a comprehensive and verifiable comparative evaluation of these two approaches versus the conventional (w.r.t. \miniKanren) Robinson's algorithm taken as a baseline.
In addition, we provide verifiable certified proofs of principal algorithm properties in \Rocq proof assistant, which to our knowledge is the first mechanized justification of rational unification algorithm correctness.

The paper is organized as follows. In Section 2 we introduce rational terms, equation systems and our algorithm. In Section 3 we describe a simplified implementation of our algorithm in \OCaml given in a line-to-line comparison with the conventional one. Section 4 presents the results of evaluation;
Section 5 explores the related work and Section 6 concludes.

\section{Rational Unification}

In this section we introduce rational terms, equation systems and the unification algorithm. The algorithm itself is inspired by the Martelli-Rossi algorithm ``UNIFY-1''~\cite{martelli1984efficient}: it is specialized to work with a threading intermediate state (equation system) and a pair of terms instead of a number of equations, and formulated as a recursive function, so the representation of an intermediate state simplified and some steps are merged together.

\subsection{Rational Terms}

In relational (and logic) programming, we conventionally consider terms as finite expressions of the following syntax (also named Herbrand trees):
\begin{equation*}
    \begin{array}{ccl}
        \V & \text{---} & \text{variables}, \\
        \C & \text{---} & \text{constructors}, \\
        T & ::= & \mathbb{V} \mid \mathbb{C}\,(T, \dots, T). \\
    \end{array}
\end{equation*}

This notion could be generalized to consider also possibly-infinite expressions of the same syntax which we name infinite terms. While ``truly'' infinite terms like $f_1\,(f_2\,(f_3\,(\dots)))$ are not much semantically helpful in computational logic (since we cannot run programs infinitely long), it is pretty interesting to consider rational terms.

Rational term is an infinite term with the only finite set of subterms~\cite{colmerauer1982prolog} (which are subtrees). For example, $f\,(f\,(f\,(\dots)))$ is a rational term since it have only one subterm (itself).

Formally, we consider countable infinite sets $x, y, z, ... \in \V$ of variables and $f, g, h, ... \in \C$ of constructors. The latter is equipped with the function $\arity : \C \rightarrow \N$ which determines the arities of constructors and the notation $f^n \in \C$ where $f \in \C$ and $n = \arity(f)$.

We consider an endofunctor $F : \text{Set} \rightarrow \text{Set}$ and define the sets of terms $T = \mu F$ and infinite terms $T^\infty = \nu F$ as the least fixed point and the greatest fixed point~\cite{simon2006coinductive} of $F$ respectively:
\[F(X) = \V \cup \{ f^n\,(t_1, \dots, t_n) \mid f^n \in \C, t_i \in X \}.\]

The set of subterms of a term, $\subterms : T^\infty \rightarrow 2^{T^\infty}$, is defined in a conventional way inductively:
\begin{itemize}
    \item $t \in \subterms(t)$;
    \item $t \in \subterms(t_i) \implies t \in \subterms(f^n\,(\dots, t_i, \dots))$.
\end{itemize}

Then, the set of rational terms is defined as $T^R = \{ t \in T^\infty \mid \subterms(t) \text{ is finite} \}$. Thus, $T \subset T^R \subset T^\infty$.

We define a substitution $\sigma \in \Sigma$ as a mapping $\V \rightarrow T^\infty$ with a finite substitution domain $\dom(\sigma) = \{ x \in \V \mid \sigma(x) \neq x \}$. Substitution application and composition operations, generality and unifier properties are defined conventionally~\cite{baader2001unification}:
\begin{itemize}
    \item Substitution application is denoted as a postfix form: $t \sigma$;
    \item Substitution composition ``$\diamond$'' respects the equality $t (\sigma_1 \diamond \sigma_2) = t\sigma_2\sigma_1$;
    \item ``More general'' relation ``$\preceq$'' is defined as follows: $\sigma_1 \preceq \sigma_2 \iff \exists \sigma, \sigma_2 = \sigma \diamond \sigma_1$.
\end{itemize}

Also, we consider a generalization of the most general unifier (MGU) property named ``the minimal unifying extension''. We say that $\sigma_2$ is the minimal unifying extension of $\sigma_1$ for terms $t_1, t_2 \in T^\infty$ iff:
\[ \sigma_1 \preceq \sigma_2 \land t_1 \sigma_2 = t_2 \sigma_2 \land \forall \sigma \in \Sigma. ~ \sigma_1 \preceq \sigma \land t_1 \sigma = t_2 \sigma \implies \sigma_2 \preceq \sigma. \]

Thus, $\sigma$ is the MGU for terms $t_1, t_2$ iff it is the minimal unifying extension of the empty substitution.

\subsection{Equation Systems}

For convenience, we will use the notation $A^? = \{ \bot \} \cup A$ for some set $A$ and call it ``optional $A$''.

We define an equation $e \in \E$ as a pair:
\begin{itemize}
    \item a pointed finite set $X \subset \V$ named ``left-hand side'' with some variable named ``representing variable'' chosen as a base point;
    \item an optional non-variable term $t \in T^? \setminus \V$ named ``right-hand side''.
\end{itemize}

For convenience, we will use the notations $\lhs : \E \rightarrow 2^\V$, $\repr : \E \rightarrow \V$ and $\rhs : \E \rightarrow T^?$ for corresponding equation components. Also, we will use the notation $\underline{x}, y, z \equiv t$, where $\lhs(\underline{x}, y, z \equiv t) = \{ x, y, z \}$, $\repr(\underline{x}, y, z \equiv t) = x$ and $\rhs(\underline{x}, y, z \equiv t) = t$.

We define an equation system $\xi \in \Xi$ as a finite set of equations $\Xi \subset 2^\E$ with pairwise disjunctive left-hand sides. For convenience, we will use the notation $\xi, e$ for the equation system extension $\{ e \} \cup \{ e' \in \xi \mid \lhs(e) \cap \lhs(e') = \emptyset \}$.

We consider our equation definition as a representation of the special case of equation system in conventional sense. For example:
\begin{equation*}
    \begin{array}{c|c}
        \text{Conventional equation system} & \text{Our equation} \\[2pt]
        \left\{\begin{array}{l@{{}={}}l}
            x & f\,(z) \\
            y & x \\
            z & x \\
        \end{array}\right. & \underline{x}, y, z \equiv f\,(z) \\
    \end{array}
\end{equation*}

In this sense, trivial equations of the form $\underline{x} \equiv \bot$ do not equate anything so we consider them illegal, i.e., $\forall x \in \V, x \in \Xi. ~ x \equiv \bot \notin \xi$. However, we will still allow them in $\E$ for the simplicity of the following definitions.

Also, under this restriction an equation system becomes a variant of the union-find~\cite{arden1961algorithm} data structure over variables where each equivalence class is attributed with an optional term. To abstract the theoretical definition from the implementation details, we are presenting it in a pure set-theoretic terms.

The main operation defined on equation systems is $\walk : \Xi \times \V \rightarrow \E \times T$. This operation looks up for a variable in an equation system returning a corresponding pair of an equation and a term. More formally:

\begin{equation*}
    \walk(\xi, x) = \begin{cases}
        (\underline{x} \equiv \bot, x) & \text{if there is no $e \in \xi$ such that $x \in \lhs(e)$}, \\
        (e, \repr(e)) & \text{if $e \in \xi, x \in \lhs(e)$ and $\rhs(e) = \bot$}, \\
        (e, \rhs(e)) & \text{otherwise}. \\
    \end{cases}
\end{equation*}

Basically, we may consider our definition of equation system as an ``enhanced'' triangular substitution~\cite{baader2001unification} w.r.t. \walk. For example:
\begin{equation*}
    \begin{array}{c|c}
        \text{Triangular substitution} & \text{Equation system} \\[2pt]
        \left\{\begin{array}{l@{{}\mapsto{}}l}
            x & f\,(z) \\
            y & x \\
            z & w \\
        \end{array}\right. & \left\{\begin{array}{l@{{}\equiv{}}l}
            \underline{x}, y & f\,(z) \\
            z, \underline{w} & \bot \\
        \end{array}\right.
    \end{array}
\end{equation*}

Indeed, we may consider equation systems as substitutions which are being applied recursively to a term. More formally, we define an equation system application as follows:
\begin{algorithmic}
    \Function{ApplyEqSys}{$\xi \in \Xi$, $t \in T^\infty$}
        \If{$x \gets t \in \V$}
            \State $(\_, t) \gets \walk(\xi, x)$
        \EndIf
        \If{$x' \gets t \in \V$}
            \State \Return $x'$
            \Comment{reached the fixed point}
        \EndIf
        \If{$f^n\,(t_1, \dots, t_n) \gets t$}
            \State \Return $f^n\,(\Call{ApplyEqSys}{\xi, t_1}, \dots, \Call{ApplyEqSys}{\xi, t_n})$
        \EndIf
    \EndFunction
\end{algorithmic}

Under the equation system application we may consider equation systems as a special case of substitutions. More precisely, we may prove that the application of an equation system to a rational term is always rational, too.

The second key operation of equation systems is $\union : \Xi \times \V \times \V \rightarrow \Xi \times (T \times T)^?$. Given two variables $x, y$ and an equation system $\xi$, this operation builds a new equation system from $\xi$ such that $x, y$ belong to the same equation. More formally:
\begin{algorithmic}[1]
    \Function{UnionEqSys}{$\xi \in \Xi$, $x, y \in \V$}
        \State $(e_x, \_) \gets \walk(\xi, x)$
        \State $(e_y, \_) \gets \walk(\xi, y)$
        \If{$\repr(e_x) = \repr(e_y)$}
            \State \Return $(\xi, \bot)$
            \Comment{already in the same equation or equals}
        \EndIf
        \If{$\rhs(e_x) = \bot$ \textbf{and} $\rhs(e_y) = \bot$}
            \State \Return $([\xi, \lhs(e_x) \cup \lhs(e_y) \equiv \bot], \bot)$
        \EndIf
        \If{$\rhs(e_x) = \bot$}
            \State \Return $([\xi, \lhs(e_x) \cup \lhs(e_y) \equiv \rhs(e_y)], \bot)$
        \EndIf
        \If{$\rhs(e_y) = \bot$}
            \State \Return $([\xi, \lhs(e_x) \cup \lhs(e_y) \equiv \rhs(e_x)], \bot)$
        \EndIf
        \State \Return $([\xi, \lhs(e_x) \cup \lhs(e_y) \equiv \rhs(e_x)],(\rhs(e_x), \rhs(e_y)))$
        \Comment{conflict in the right-hand sides}
    \EndFunction
\end{algorithmic}

When we add a new equation at lines 6-12, we may choose either $\repr(e_x)$ or $\repr(e_y)$ as the representing variable. The same is true about the right-hand side of the new equation at line 12: it could be either $\rhs(e_x)$ or $\rhs(e_y)$. As the listing shows, the second element of \union result is intended to inform a caller about a collision in right-hand sides during the union.

\subsection{Common Part}

We define a ``common part''~\cite{martelli1984efficient} of two terms $\commonpart : T \times T \rightarrow T^?$ inductively:
\begin{equation*}
    \begin{split}
        \commonpart(x, \_) &= x \\
        \commonpart(\_, y) &= y \\
        \commonpart(f^n\,(l_1, \dots, l_n), f^n\,(r_1, \dots, r_n)) &= f^n\,(\commonpart(l_1, r_1), \dots, \commonpart(l_n, r_n)) \\
    \end{split}
\end{equation*}

The definition is intentionally left non-total since two terms may not have a common part (in such cases the result is assumed to be $\bot$) and slightly non-deterministic (in the case of two variables) to allow optimizations. Hereafter we require the implementation of \commonpart to be a total deterministic function.

\subsection{The Unification Algorithm}

We define the unification algorithm as a recursive function $\unify : \Xi \times T \times T \rightarrow \Xi^?$:

\begin{algorithmic}[1]
    \Function{RationalUnify}{$\xi \in \Xi$, $t_1, t_2 \in T$}
        \If{$x \gets t_1 \in \V$ \textbf{and} $y \gets t_2 \in \V$}
            \State $(\xi, ts) \gets \union(\xi, x, y)$
            \Comment{union the equations}
            \If{$(t_1, t_2) \gets ts$}
                \Comment{check for a collision and resolve it}
                \State \Return $\Call{RationalUnify}{\xi, t_1, t_2}$
            \EndIf
            \State \Return $\xi$
        \EndIf
        \If{$x \gets t_1 \in \V$}
            \State $(e, \_) \gets \walk(\xi, x)$
            \If{$\rhs(e) = \bot$}
                \State \Return $\lhs(e) \equiv t_2, \xi$
                \Comment{bind the unbound variable}
            \EndIf
            \State $t \gets \commonpart(\rhs(e), t_2)$
            \If{$t = \bot$}
                \State {\bf failure}
                \Comment{there is no common part}
            \EndIf
            \State $\xi \gets \lhs(e) \equiv t, \xi$
            \Comment{simplifying the right-hand side}
            \State \Return \Call{RationalUnify}{$\xi, \rhs(e), t_2$}
            \Comment{handling the residual parts of the terms}
        \EndIf
        \If{$y \gets t_2 \in \V$}
            \State \Return \Call{RationalUnify}{$\xi, y, t_1$}
        \EndIf
        \If{$f^n\,(l_1, ..., l_n) \gets t_1$ \textbf{and} $g^m\,(r_1, ..., r_m) \gets t_2$}
            \If{$f^n \neq g^m$}
                \State \bf failure
            \EndIf
            \For{$i = 1, ..., n$}
                \State $\xi \gets \Call{RationalUnify}{\xi, l_i, r_i}$
            \EndFor
            \State \Return $\xi$
        \EndIf
    \EndFunction
\end{algorithmic}

We have proved the following properties of this algorithm in \Rocq\footnote{\url{https://github.com/dboulytchev/miniKanren-coq/tree/master/src/Rational}}:

\begin{itemize}
    \item Soundness: if the algorithm terminates with a non-$\bot$ result then the result is the minimal unifying extension of the source equation system for given terms.
    \item Completeness: if the algorithm terminates with result $\bot$ then there is no unifying extension of the source equation system for given terms.
    \item Termination: the algorithm terminates with some result on every input.
\end{itemize}

While soundness and completeness of the algorithm are mostly intuitive, its termination is a bit unclear. We consider two characteristics of an equation system: the number of ``roots'' and the number of constructors in right-hand sides (``size''). Intuitively, a root is an equivalence class of variables w.r.t. an equation system: every equation represents one by its left-hand side. Also, we consider a ``cost'' of two terms which is a number of their constructors, counting shared ones (w.r.t. \commonpart) only once.

We have proved that any run $(\Xi, t_1, t_2)$ of the algorithm is performed on a decreasing sequence of pairs $(\textsf{roots}(\Xi), \textsf{size}(\Xi) + \textsf{cost}(t_1, t_2))$
compared lexicographically, so it won't loop infinitely. The details are outside of the scope of the paper; an interested reader can consult the repository with the \Rocq implementation.

We omit the proof because it would require describing the algorithm ``UNIFY-1'' of Martelli and Rossi~\cite{martelli1984efficient}. Any execution of our algorithm can be simulated by a sequence of steps of ``UNIFY-1'' w.r.t. the intermediate state changes. Their complexity analysis yields the asymptotic bound $\mathcal{O}(n + m \alpha(m))$ where $n$ is the number of constructor occurrences, $m$ is the number of variable occurrences and $\alpha(\cdot)$ is the inverse of the Ackermann function. Consequently, our algorithm has the same asymptotic complexity where $n$ and $m$ are measured w.r.t. both the equation system and the pair of input terms. Indeed, this bound is primarily of theoretical interest, since practical implementations often omit path compression and the union-by-rank heuristic for performance reasons. For a persistent implementation, the second term may incur an additional logarithmic factor in the conventional BST-based implementation, or take a different asymptotic form depending on the underlying persistent data structure (e.g.~\cite{conchon2007persistent}).

\subsection{Examples}

To illustrate the work of the algorithm we consider here some concrete examples comparing intermediate states of the conventional one and ours. To give more intuition
about the termination we are using a variant of the conventional algorithm without ``occurs check''.

Starting from the empty state, let us unify the variable $x$ with the term $f\,(f\,(x))$. Both algorithms are observing that $x$ is unbound and stop immediately:
\begin{equation*}
    \begin{array}{c|c}
        \text{Triangular substitution} & \text{Equation system} \\[2pt]
        \left\{\begin{array}{l@{{}\mapsto{}}l}
            x & f\,(f\,(x)) \\
        \end{array}\right. & \left\{\begin{array}{l@{{}\equiv{}}l}
            \underline{x} & f\,(f\,(x)) \\
        \end{array}\right.
    \end{array}
\end{equation*}

Next, we are to unify the variable $y$ with the term $f\,(f\,(y))$ in the same manner:
\begin{equation}
    \begin{array}{c|c}
        \text{Triangular substitution} & \text{Equation system} \\[2pt]
        \left\{\begin{array}{l@{{}\mapsto{}}l}
            x & f\,(f\,(x)) \\
            y & f\,(f\,(y)) \\
        \end{array}\right. & \left\{\begin{array}{l@{{}\equiv{}}l}
            \underline{x} & f\,(f\,(x)) \\
            \underline{y} & f\,(f\,(y)) \\
        \end{array}\right.
    \end{array}
    \label{eq:example-state}
\end{equation}

To demonstrate the role of \union, let us trace the unification of the variables $x$ and $y$:
\begin{center}
    \begin{longtable}{p{4cm}|p{7cm}}
        {\centering Conventional \par} & {\centering Our \par} \\
        \hline\hline\endhead
        {
            \textbf{Input: $x, y$}
            \begin{dinglist}{212}
                \item $\walk(x) = f\,(f\,(x))$
                \item $\walk(y) = f\,(f\,(y))$
            \end{dinglist}
        } & {
            \textbf{Input: $x, y$}
            \begin{dinglist}{212}
                \item $\union(x, y) = (\dots, (f\,(f\,(x)), f\,(f\,(y))))$
                \item Update equation system: \newline
                \begin{minipage}{\linewidth}
                    \begin{equation*}
                        \left\{\begin{array}{l@{{}\equiv{}}l}
                            \underline{x}, y & f\,(f\,(x)) \\
                        \end{array}\right.
                    \end{equation*}
                \end{minipage}
            \end{dinglist}
        } \\\\\hline
        {
            \textbf{Input: $f\,(f\,(x)), f\,(f\,(y))$}
            \begin{dinglist}{212}
                \item Unwrap $f\,(\cdot)$
            \end{dinglist}
        } & {
            \textbf{Input: $f\,(f\,(x)), f\,(f\,(y))$}
            \begin{dinglist}{212}
                \item Unwrap $f\,(\cdot)$
            \end{dinglist}
        } \\\hline
        {
            \textbf{Input: $f\,(x), f\,(y)$}
            \begin{dinglist}{212}
                \item Unwrap $f\,(\cdot)$
            \end{dinglist}
        } & {
            \textbf{Input: $f\,(x), f\,(y)$}
            \begin{dinglist}{212}
                \item Unwrap $f\,(\cdot)$
            \end{dinglist}
        } \\\hline
        {
            \textbf{Input: $x, y$}
            \begin{dinglist}{212}
                \item Loops infinitely...
            \end{dinglist}
        } & {
            \textbf{Input: $x, y$}
            \begin{dinglist}{212}
                \item $\union(x, y) = (\text{same equation system}, \bot)$
            \end{dinglist}
        } \\
    \end{longtable}
\end{center}

This example uncovers how \textit{proactively} updating the intermediate state avoids infinite looping in simple cases where the term cycles directly on a variable.

Similarly, let us undo this (returning to the state marked~\ref{eq:example-state}) to demonstrate the contribution of \commonpart, tracing the unification of the variable $x$ and the term $f\,(y)$:
\begin{center}
    \begin{longtable}{p{4cm}|p{6.5cm}}
        {\centering Conventional \par} & {\centering Our \par} \\
        \hline\hline\endhead
        {
            \textbf{Input: $x, f\,(y)$}
            \begin{dinglist}{212}
                \item $\walk(x) = f\,(f\,(x))$
            \end{dinglist}
        } & {
            \textbf{Input: $x, f\,(y)$}
            \begin{dinglist}{212}
                \item $\walk(x) = (\underline{x} \equiv f\,(f\,(x)), f\,(f\,(x)))$
                \item $\commonpart(f\,(f\,(x)), f\,(y)) = f\,(y)$
                \item Update equation system: \newline
                \begin{minipage}{\linewidth}
                    \begin{equation*}
                        \left\{\begin{array}{l@{{}\equiv{}}l}
                            \underline{x} & f\,(y) \\
                            \underline{y} & f\,(f\,(y)) \\
                        \end{array}\right.
                    \end{equation*}
                \end{minipage}
            \end{dinglist}
        } \\\\\hline
        {
            \textbf{Input: $f\,(f\,(x)), f\,(y)$}
            \begin{dinglist}{212}
                \item Unwrap $f\,(\cdot)$
            \end{dinglist}
        } & {
            \textbf{Input: $f\,(f\,(x)), f\,(y)$}
            \begin{dinglist}{212}
                \item Unwrap $f\,(\cdot)$
            \end{dinglist}
        } \\\hline
        {
            \textbf{Input: $f\,(x), y$}
            \begin{dinglist}{212}
                \item $\walk(y) = f\,(f\,(y))$
            \end{dinglist}
        } & {
            \textbf{Input: $f\,(x), y$}
            \begin{dinglist}{212}
                \item $\walk(y) = (\underline{y} \equiv f\,(f\,(y)), f\,(f\,(y)))$
                \item $\commonpart(f\,(x), f\,(f\,(y))) = f\,(x)$
                \item Update equation system: \newline
                \begin{minipage}{\linewidth}
                    \begin{equation*}
                        \left\{\begin{array}{l@{{}\equiv{}}l}
                            \underline{x} & f\,(y) \\
                            \underline{y} & f\,(x) \\
                        \end{array}\right.
                    \end{equation*}
                \end{minipage}
            \end{dinglist}
        } \\\\\hline
        {
            \textbf{Input: $f\,(x), f\,(f\,(y))$}
            \begin{dinglist}{212}
                \item Unwrap $f\,(\cdot)$
            \end{dinglist}
        } & {
            \textbf{Input: $f\,(x), f\,(f\,(y))$}
            \begin{dinglist}{212}
                \item Unwrap $f\,(\cdot)$
            \end{dinglist}
        } \\\hline
        {
            \textbf{Input: $x, f\,(y)$}
            \begin{dinglist}{212}
                \item Loops infinitely...
            \end{dinglist}
        } & {
            \textbf{Input: $x, f\,(y)$}
            \begin{dinglist}{212}
                \item $\walk(x) = (\underline{x} \equiv f\,(y), f\,(y))$
                \item $\commonpart(f\,(y), f\,(y)) = f\,(y)$
                \item Update equation system (no-op)
            \end{dinglist}
        } \\\cline{2-2}
        & {
            \textbf{Input: $f\,(y), f\,(y)$}
            \begin{dinglist}{212}
                \item Unwrap $f\,(\cdot)$
            \end{dinglist}
        } \\\cline{2-2}
        & {
            \textbf{Input: $y, y$}
            \begin{dinglist}{212}
                \item Do \union (no-op)
            \end{dinglist}
        } \\
    \end{longtable}
\end{center}

As the example shows, \commonpart plays a crucial role in termination gradually simplifying the intermediate state until reaching the fixed point. This allows the algorithm to detect cycling terms in more complicated cases when the initial state doesn't include a cycling part itself.

\section{Implementation}

In this section we present the implementation of our algorithm and compare it with the conventional unification algorithm used in the majority of \miniKanren implementations
(which is in turn based on Robinson's algorithm with triangular substitutions~\cite{baader2001unification}). Specifically, we consider a simplified version of \OCanren
unification implementation.

In both algorithms the datatype for the state is a mapping from variables to terms:

\begin{minted}{OCaml}
module Term = struct
  type t =
  | Var of int
  | Con of int * t array
end
module M = Map.Make(Int)
\end{minted}

The difference is in the interpretation of this intermediate state. While in conventional unification we understand an intermediate state as a substitution $\sigma$ (encoded
in a triangular manner~\cite{baader2001unification}), in rational unification we understand it as an equation system $\xi$ encoded in a union-find manner~\cite{arden1961algorithm}\footnote{Actually, we don't use path compression and the union-by-rank heuristic since they slow down the algorithm in persistent implementations. Thus, it may be called ``a triangular equation system''.}.

The core operation \walk which allows us to ``dereference'' variables in triangular substitution changes its signature from $\walk : \Sigma \times \V \rightarrow T$ to $\walk : \Xi \times \V \rightarrow \E \times T$ and plays the role of the ``find'' operation of union-find:

\begin{center}
    \begin{minipage}{0.46\textwidth}
        \centering
        \begin{minted}[fontsize=\small,baselinestretch=0.9]{OCaml}
let rec walk s x = match M.find x s with
| exception Not_found -> Term.Var x
| Term.Var x -> walk s x
| t -> t
        \end{minted}
    \end{minipage}
    \hfill
    \vrule
    \hfill
    \begin{minipage}{0.46\textwidth}
        \centering
        \begin{minted}[fontsize=\small,baselinestretch=0.9]{OCaml}
let rec walk s x = match M.find x s with
| exception Not_found -> x, Term.Var x
| Term.Var x -> walk s x
| t -> x, t
        \end{minted}
    \end{minipage}
\end{center}

The only difference is an ability to observe the ``final'' variable in a ``dereference'' chain. In implementation, we deal with representing variables instead of equations defined in the previous section.
\pagebreak

Let us recall the ``occurs check'' from the conventional algorithm:

\begin{minted}{OCaml}
let rec occurs s x = function
| Term.Var y ->
  begin match walk s y with
  | Term.Var y -> if x == y then raise Occurs
  | t -> occurs x t
  end
| Term.Con (_, ts) -> Array.iter (occurs s x) ts

let extend s x t = occurs s x t ; M.add x t s
\end{minted}
\vspace{\baselineskip}

While in the conventional algorithm these functions help us to prevent the construction of illegal (recursive) substitution, rational unification don't rely on them.

Instead, we define the operations $\commonpart : T \times T \rightarrow T^?$ and $\union : \Xi \times \V \times \V \rightarrow \Xi \times (T \times T)^?$:

\begin{minted}{OCaml}
module Term = struct
  let rec common_part x y = match x, y with
  | Var x, _ -> Var x
  | _, Var y -> Var y
  | Con (f, ts1), Con (g, ts2)
  when f == g && Array.length ts1 == Array.length ts2 ->
    Con (f, Array.map2 common_part ts1 ts2)
  | _ -> raise No_common_part
end

let union s x y =
  let x, xt = walk s x in
  let y, yt = walk s y in
  if x == y then s, None
  else begin match xt, yt with
  | Var _, Var _ -> M.add x yt s, None
  | Var _, _ -> M.add x (Term.Var y) s, None
  | _, Var _ -> M.add y (Term.Var x) s, None
  | _, _ -> M.add x (Term.Var y) s, Some (xt, yt)
  end
\end{minted}
\vspace{\baselineskip}

Finally, we implement the algorithm itself:

\begin{center}
    \setlength{\LTpre}{4pt}
    \setlength{\LTleft}{0pt}
    \setlength{\LTright}{0pt}
    \renewcommand{\arraystretch}{0}
    \setlength{\extrarowheight}{0pt}
    \begin{longtable}{@{}p{.46\textwidth}@{\extracolsep{\fill}}c|p{.46\textwidth}@{}}
        \begin{minted}[fontsize=\small,baselinestretch=0.9]{OCaml}
let rec unify_vt s x yt =
        \end{minted}
        &&
        \begin{minted}[fontsize=\small,baselinestretch=0.9]{OCaml}
let rec unify_vt s x yt =
  let x, xt = walk s x in
        \end{minted}
        \\
        \begin{minted}[fontsize=\small,baselinestretch=0.9]{OCaml}
  match walk s x with
  | Term.Var x -> extend s x yt
  | xt -> unify xt yt
        \end{minted}
        &&
        \begin{minted}[fontsize=\small,baselinestretch=0.9]{OCaml}
  begin match xt with
  | Term.Var _ -> M.add x yt s
  | _ ->
    let t = Term.common_part xt yt in
    unify (M.add x t s) xt yt
  end
        \end{minted}
        \\
        \begin{minted}[fontsize=\small,baselinestretch=0.9]{OCaml}
and unify s xt yt =
  if xt == yt then s
  else match xt, yt with
        \end{minted}
        &&
        \begin{minted}[fontsize=\small,baselinestretch=0.9]{OCaml}
and unify s xt yt =
  if xt == yt then s
  else match xt, yt with
        \end{minted}
        \\
        \begin{minted}[fontsize=\small,baselinestretch=0.9]{OCaml}
  | Term.Var x, Term.Var y ->
    begin match walk s x, walk s y with
    | Term.Var x, Term.Var y ->
      M.add x (Term.Var y) s
    | Term.Var x, yt -> extend s x yt
    | xt, Term.Var y -> extend s y xt
    | xt, yt -> unify s xt yt
    end
        \end{minted}
        &&
        \begin{minted}[fontsize=\small,baselinestretch=0.9]{OCaml}
  | Term.Var x, Term.Var y ->
    begin match union s x y with
    | s, Some (xt, yt) -> unify s xt yt
    | s, None -> s
    end
        \end{minted}
        \\
        \begin{minted}[fontsize=\small,baselinestretch=0.9]{OCaml}
  | Term.Var x, _ -> unify_vt s x yt
  | _, Term.Var y -> unify_vt s y xt
  | Term.Con (f, ts1), Term.Con (g, ts2)
  when f == g && Array.length ts1
              == Array.length ts2 ->
    let n = Array.length ts1 in
    let rec hlp i s =
      if i == n then s
      else hlp (i + 1)
        @@ unify s ts1.(i) ts2.(i)
    in
    hlp 0 s
  | _ -> raise Term.No_common_part
        \end{minted}
        &&
        \begin{minted}[fontsize=\small,baselinestretch=0.9]{OCaml}
  | Term.Var x, _ -> unify_vt s x yt
  | _, Term.Var y -> unify_vt s y xt
  | Term.Con (f, ts1), Term.Con (g, ts2)
  when f == g && Array.length ts1
              == Array.length ts2 ->
    let n = Array.length ts1 in
    let rec hlp i s =
      if i == n then s
      else hlp (i + 1)
        @@ unify s ts1.(i) ts2.(i)
    in
    hlp 0 s
  | _ -> raise Term.No_common_part
        \end{minted}
        \\
    \end{longtable}
\end{center}

While handling the two-constructors case is the same, the changes affect other cases:

\begin{itemize}
    \item The logic of unification of two variables completely migrates in \union.
    \item The logic of unification of a variable and a constructor-term is now updates the equation system using \commonpart before proceeding with the recursive call. 
\end{itemize}

With these changes, the unification of two variables now always assigns one variable to another.
For example, if $x$ is bound to $f\,(y)$, the result of conventional unification of $x$ and $z$ includes the binding $z \mapsto f\,(y)$, while the result of our unification includes $z \mapsto x$.
Also, \union \textit{proactively} binds variables, i.e., if $x$ is bound to $f\,(v)$ and $y$ is bound to $f\,(u)$, the result of conventional unification of $x$ and $y$ includes the binding $v \mapsto u$ but not $x \mapsto y$, while the result of our unification includes both.

The use of \commonpart allows the unification algorithm to decompose ``complex'' terms in the context of the current equation system.
For example, if $x$ is bound to $f\,(g\,(y))$, the result of conventional unification of $x$ and $f\,(z)$ includes $x \mapsto f\,(g\,(y)), z \mapsto g\,(y)$, while the result of our unification includes $x \mapsto f\,(z), z \mapsto g\,(y)$ instead.

While these changes are required to prevent infinite looping and deliver the correctness of the algorithm, they abuse variables reassigning in intermediate states.
This makes it more complicated to apply scope-based optimization~\cite{byrd2009relational} widely used in \miniKanren implementations. More precisely, we need to check a substitution in a state before an attribute one~--- since it seems too wasteful, in practice, substitutions in states become noticeably small in presence of the optimization, so the performance degradation is acceptable (based on some experiments and not verified strictly).

A crucial optimization for the presented algorithm is a well-known variable-age heuristic~\cite{mannila1986complexity,domoratskiy2025empirical}, which reduces \walk paths. The heuristic is directly applicable within \union.

\subsection{Optimized Common Part}

A notable optimization is the implementation of \commonpart which preserves ``physical equality''\footnote{``Physical equality'' is the equality of the references to the objects in memory.} as long as possible. The optimization is motivated by both the reduction of memory consumption and the efficiency improvement of the unification itself (since we may do a fail-fast ``physical equality'' check).

The core idea is to consider a simple lattice of four elements (we name it ``side''): $\top$~(\texttt{Both}), $\Leftarrow$~(\texttt{Left}), $\Rightarrow$~(\texttt{Right}) and $\bot$~(\texttt{Aside}):
\begin{center}
    \begin{tikzpicture}
        \node (both) at (0, 1) {$\top$};
        \node (left) at (-1, 0) {$\Leftarrow$};
        \node (right) at (1, 0) {$\Rightarrow$};
        \node (aside) at (0, -1) {$\bot$};
        \draw[dotted] (left) -- (both) node[midway,sloped] {$<$};
        \draw[dotted] (right) -- (both) node[midway,sloped] {$>$};
        \draw[dotted] (aside) -- (left) node[midway,sloped] {$>$};
        \draw[dotted] (aside) -- (right) node[midway,sloped] {$<$};
        \node[anchor=south] at (both.north) {(\texttt{Both})};
        \node[anchor=east] at (left.west) {(\texttt{Left})};
        \node[anchor=west] at (right.east) {(\texttt{Right})};
        \node[anchor=north] at (aside.south) {(\texttt{Aside})};
    \end{tikzpicture}
\end{center}

This lattice designates which of the arguments are ``physically equal'' to their \commonpart. With it, the optimization is captured by the function $\commonpart' : T \times T \rightarrow (T \times \textsf{Side})^?$. While non-recursive cases are trivial, during recursive calls we are accumulating the infimum of the ``sides'':

\begin{itemize}
    \item if the infimum is $\bot$, we are required to return the accumulated term (as in the simple implementation)~--- there is no way to preserve a ``physical equality'' to any of the arguments;
    \item otherwise, we are discarding the accumulated term and returning one of the arguments instead.
\end{itemize}

The implementation is mostly intuitive but verbose, so we omit it and refer to the evaluation repository. Also, we haven't measured the concrete performance impact of this optimization.

\section{Evaluation}

In this section we present the evaluation results of our algorithm in comparison with the conventional one (as a baseline) and another well-known rational unification algorithm.

We have implemented three unification algorithms with some optimizations\footnote{\url{https://github.com/ProgMiner/rational-unification}}.
These algorithms were implemented in three variants: ephemeral, ephemeral with path compression and persistent.
The latter uses only persistent data structures which makes it possible to utilize them in a \miniKanren implementation, while others use mutable ones without backtracking support (in contrast with WAM~\cite{warren1983abstract}, to reduce the overhead).

To provide a fair algorithm comparison all implementations include the same generic optimizations (e.g., the variable-age one) but as many as possible algorithm-specific micro-optimizations. Specifically, we use different data structures for them.

We compare the following algorithms:

\begin{itemize}
    \item \texttt{robinson}~--- the conventional algorithm (Robinson's with triangular substitution, baseline);
    \item \texttt{rational}~--- our algorithm (described above);
    \item \texttt{huet}~--- Huet's algorithm for unification of recursive types~\cite{huet1976resolution}\footnote{The actual implementation was studied using \url{https://github.com/SWI-Prolog/swipl/blob/45e83178379f0fd6f918bd294bb361f3db7bf1d7/src/pl-prims.c\#L71}.}.
\end{itemize}

We used \emph{unification traces} for evaluation. Unification trace is a list of pairs of terms, i.e., a sequence of equations. We use three test sets:

\begin{itemize}
    \item \texttt{JGS}~--- collected from the tests from Java Generic Solver implemented in \OCanren~\cite{lozov2023relational} (9 tests, 1448 traces);
    \item \texttt{Lama}~--- collected from the tests from Lama type checker implemented in \OCanren~\cite{domoratskiy2024relational} (66 traces, 1 per test);
    \item \texttt{synth}~--- synthetic tests, generated manually (66 traces, 11 per test).
\end{itemize}

We consider JGS and Lama benchmarks as unification problems close to a ``real-world'' applications of our algorithm. While JGS includes only unification of finite terms,
several Lama tests require dealing with rational ones.

Synthetic tests are specifically devised for stress testing and parameterized with a certain value $n = 0, 10, \dots, 100$:

\begin{itemize}
    \item \texttt{full} --- equates two full binary trees of depth $n$ ($T_i, U_i \in \V$):
    \begin{equation*}
        \left\{
        \begin{array}{c@{{}={}}ll}
            T_0 & x, & x \in \V, \\
            U_0 & a, & a^0 \in \C, \\
            T_i & f(T_{i-1}, T_{i-1}), & i = 1, 2, \dots, n, \\
            U_i & f(U_{i-1}, U_{i-1}), & i = 1, 2, \dots, n, \\
            T_n & U_n. \\
        \end{array}
        \right.
    \end{equation*}
    \item \texttt{left\_right} --- equates ``leftist'' and ``rightist'' binary trees of depth $n$ ($T_i, U_i \in \V$):
    \begin{equation*}
        \left\{
        \begin{array}{c@{{}={}}ll}
            T_0 & x, \\
            U_0 & x, \\
            T_i & f(T_{i-1}, x), & i = 1, 2, \dots, n, \\
            U_i & f(x, U_{i-1}), & i = 1, 2, \dots, n, \\
            T_n & U_n. \\
        \end{array}
        \right.
    \end{equation*}
    \item $\texttt{loops}_1$ --- equates $n$ loops together:
    \begin{equation*}
        \left\{
        \begin{array}{c@{{}={}}ll}
            x_i & f(x_i), & i = 1, 2, \dots, n, \\
            x_0 & x_i, & i = 1, 2, \dots, n. \\
        \end{array}
        \right.
    \end{equation*}
    \item $\texttt{loops}_2$ --- as the previous but with $x_i = x_0$ instead of $x_0 = x_i$.
    \item $\texttt{loops}_3$ --- as $\texttt{loops}_1$ but with $x_{i-1} = x_i$ instead of $x_0 = x_i$.
    \item $\texttt{loops}_4$ --- as the previous but with $x_i = x_{i-1}$ instead of $x_{i-1} = x_i$.
\end{itemize} 

All synthetic tests except for \texttt{full} require the support for rational unification. Also, \texttt{full} being run using
the conventional algorithm consumes exponential time. Consequently, we don't use the conventional algorithm for synthetic tests.

\begin{figure}[b]
    \centering
    \begin{subfigure}{.48\linewidth}
        \includegraphics[width=\linewidth]{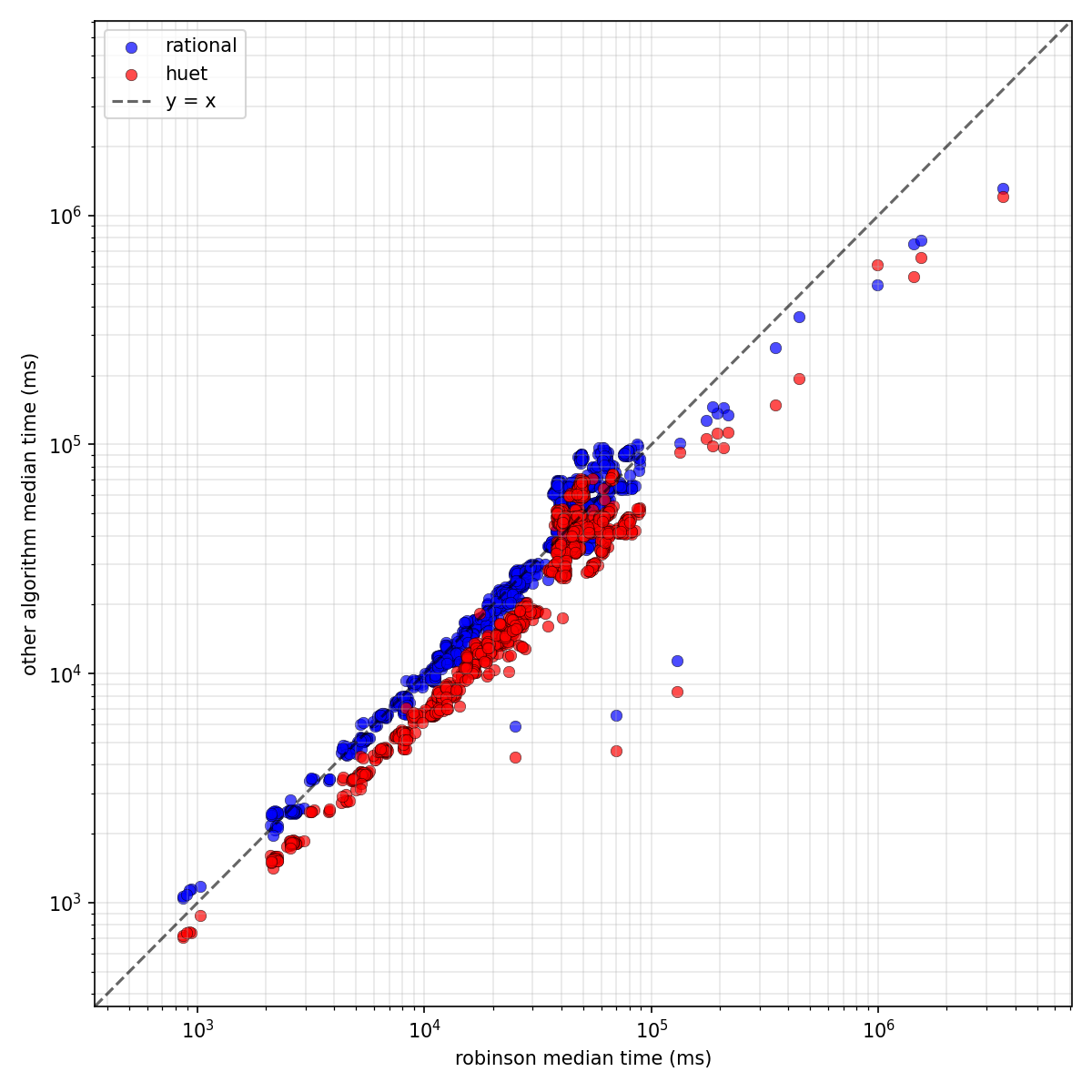}
        \subcaption{Ephemeral variant}
    \end{subfigure}
    \hfill
    \begin{subfigure}{.48\linewidth}
        \includegraphics[width=\linewidth]{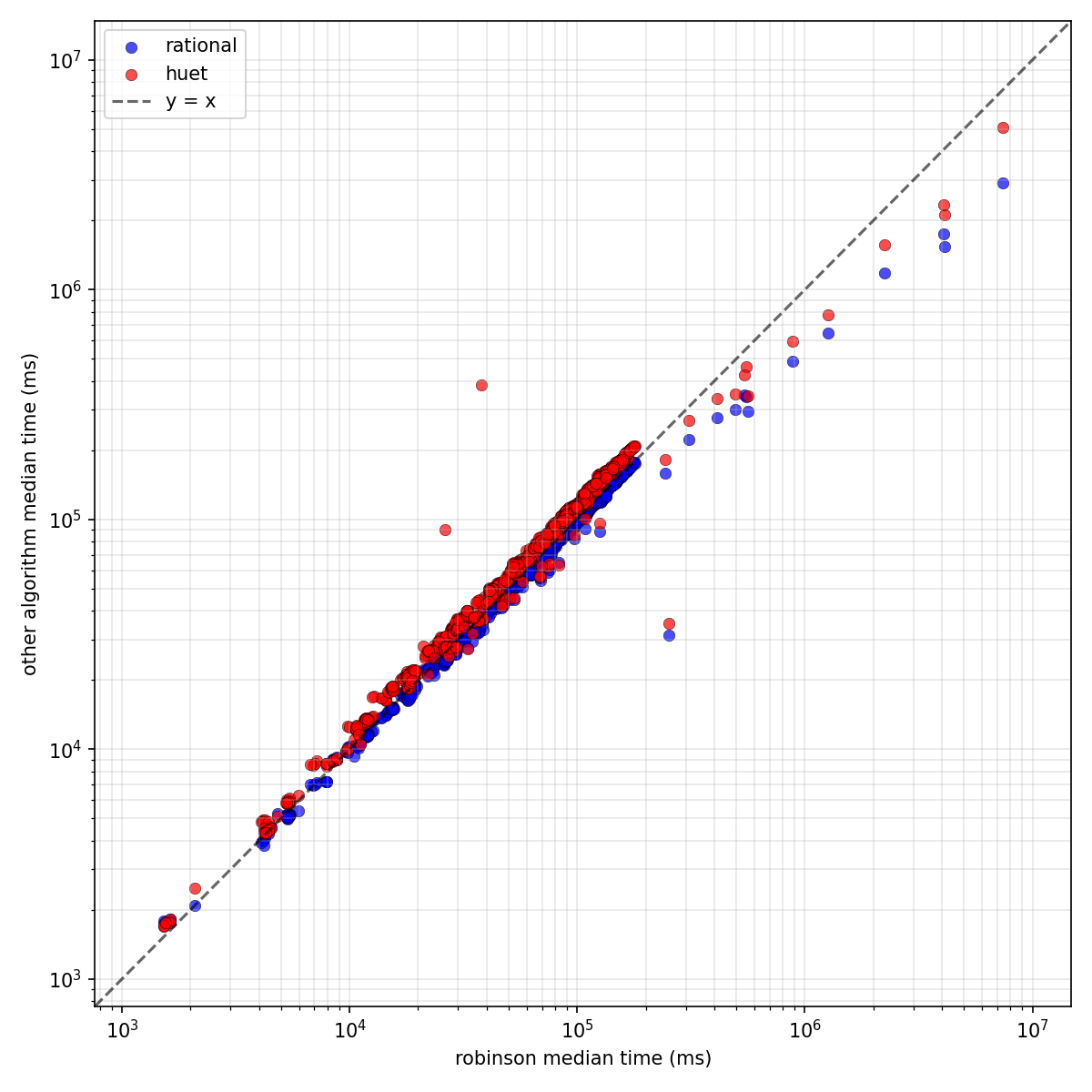}
        \subcaption{Persistent variant}
    \end{subfigure}
    \caption{Comparison of the algorithms' performance w.r.t. the conventional one}
    \label{fig:baseline}
\end{figure}

\begin{figure}
    \centering
    \begin{subfigure}{.48\linewidth}
        \includegraphics[width=\linewidth]{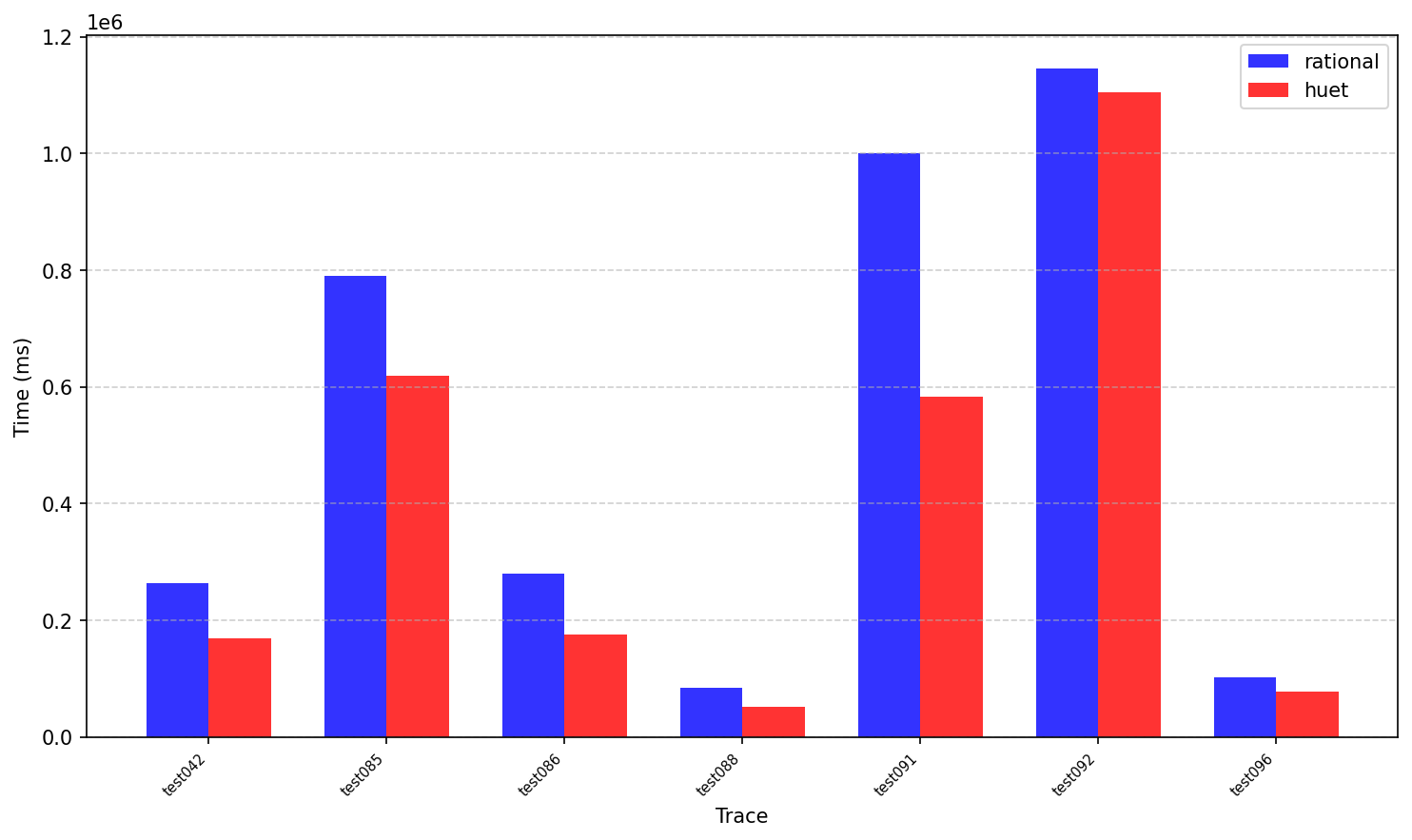}
        \subcaption{Ephemeral variant}
    \end{subfigure}
    \hfill
    \begin{subfigure}{.48\linewidth}
        \includegraphics[width=\linewidth]{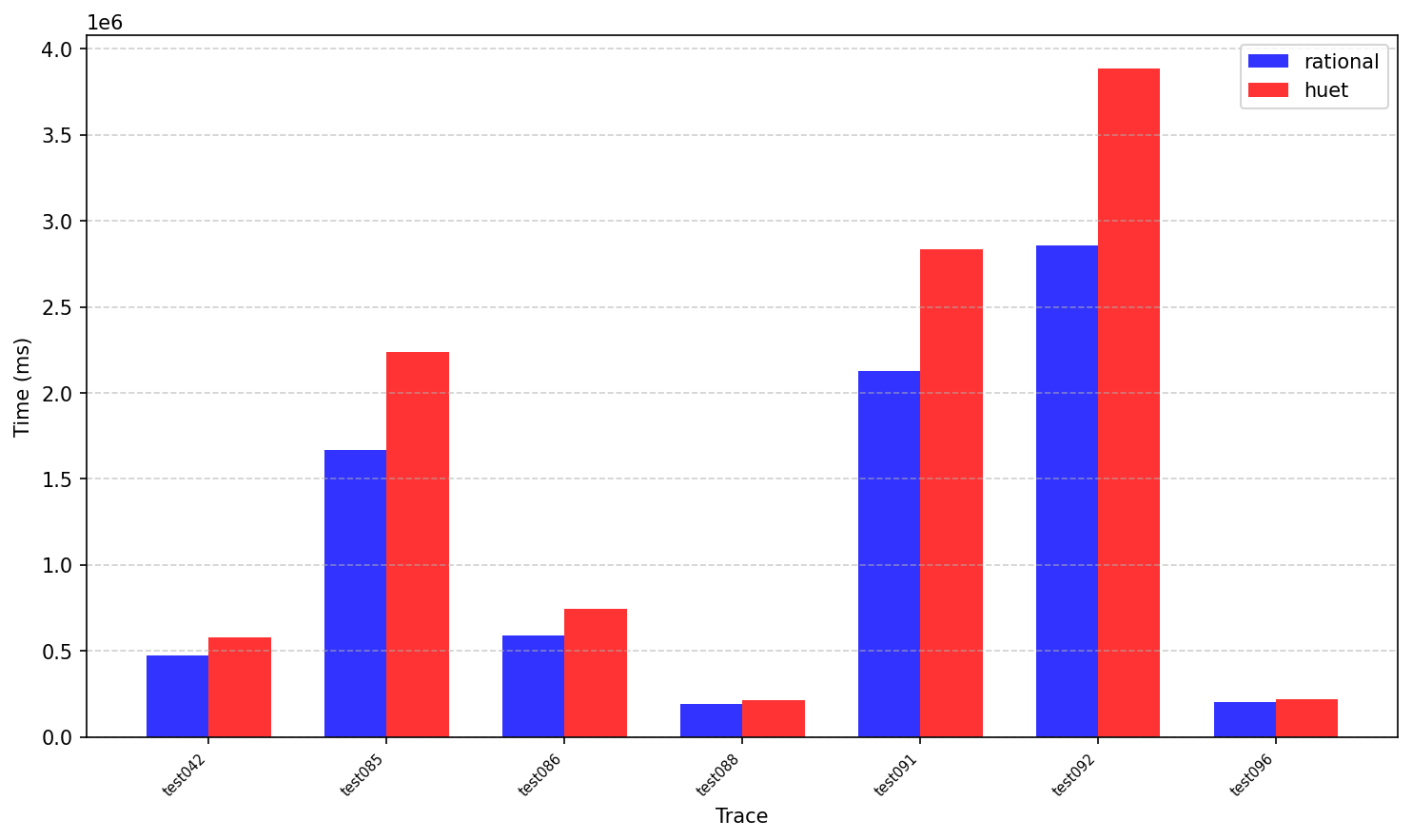}
        \subcaption{Persistent variant}
    \end{subfigure}
    \caption{Comparison of the algorithms on \texttt{Lama} traces that require rational unification}
    \label{fig:Lama}
\end{figure}

\begin{figure}
    \centering
    \begin{subfigure}{.48\linewidth}
        \includegraphics[width=\linewidth]{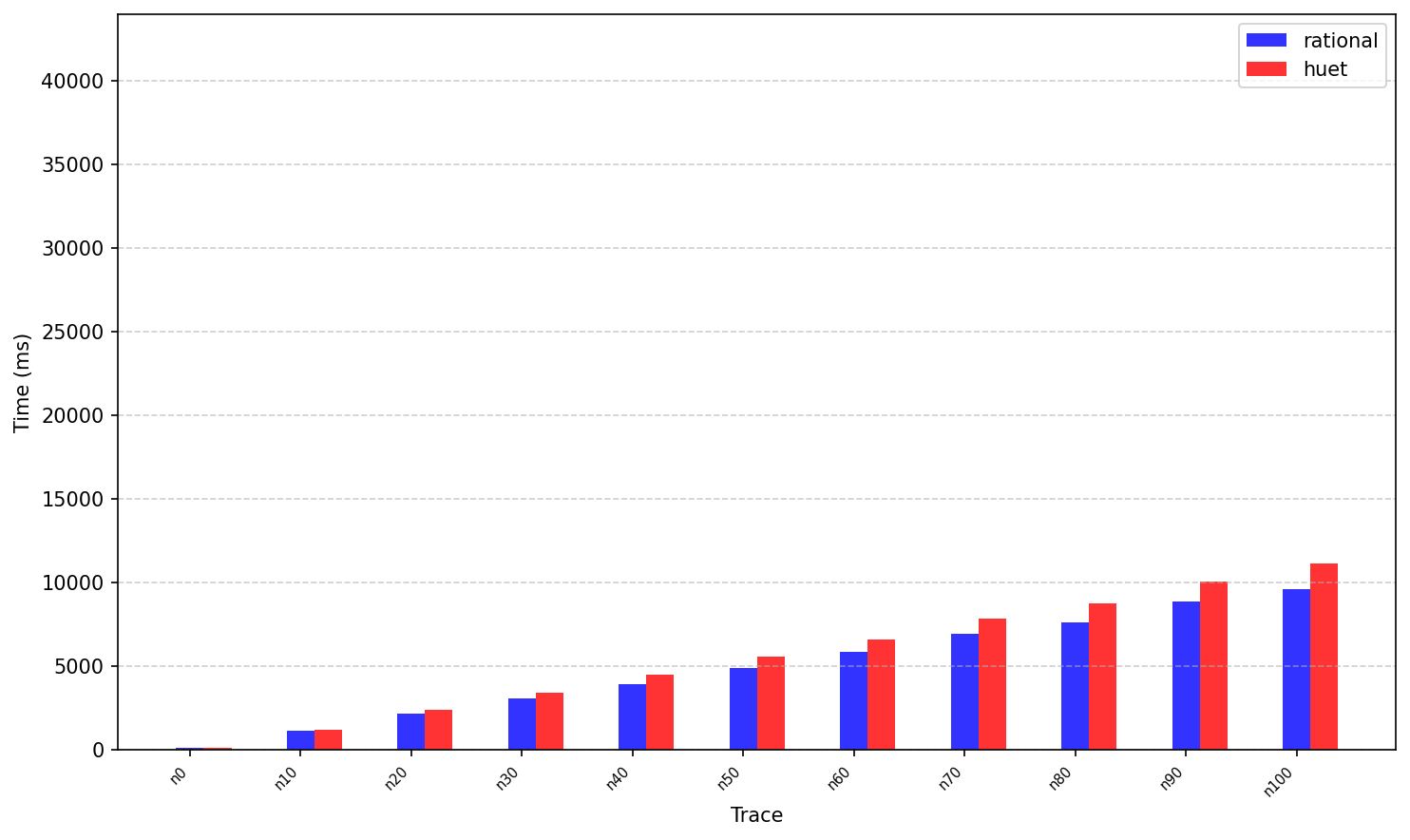}
        \subcaption{\texttt{full}}
    \end{subfigure}
    \hfill
    \begin{subfigure}{.48\linewidth}
        \includegraphics[width=\linewidth]{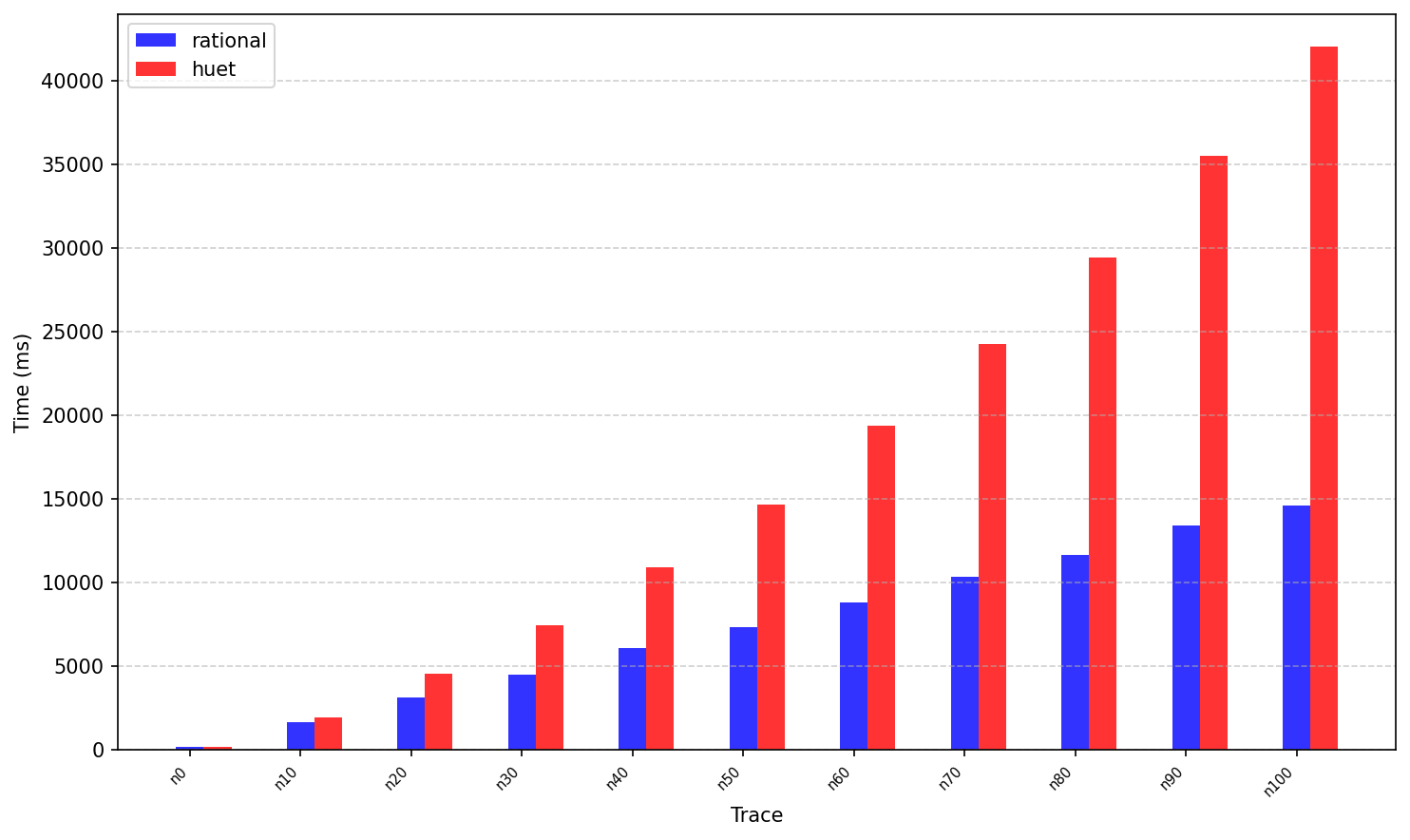}
        \subcaption{\texttt{left\_right}}
    \end{subfigure}
    \\
    \begin{subfigure}{.48\linewidth}
        \includegraphics[width=\linewidth]{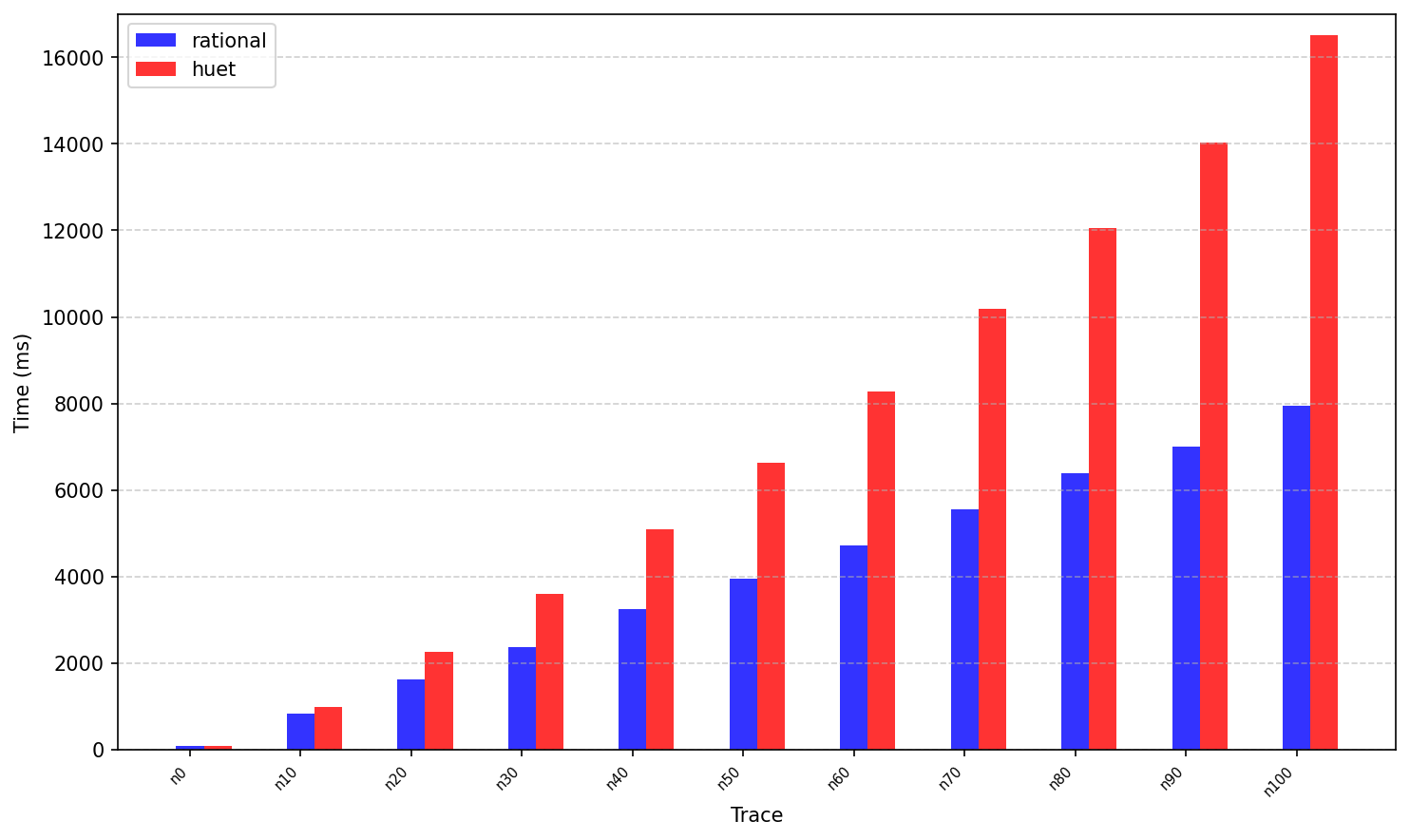}
        \subcaption{$\texttt{loops}_1$}
    \end{subfigure}
    \hfill
    \begin{subfigure}{.48\linewidth}
        \includegraphics[width=\linewidth]{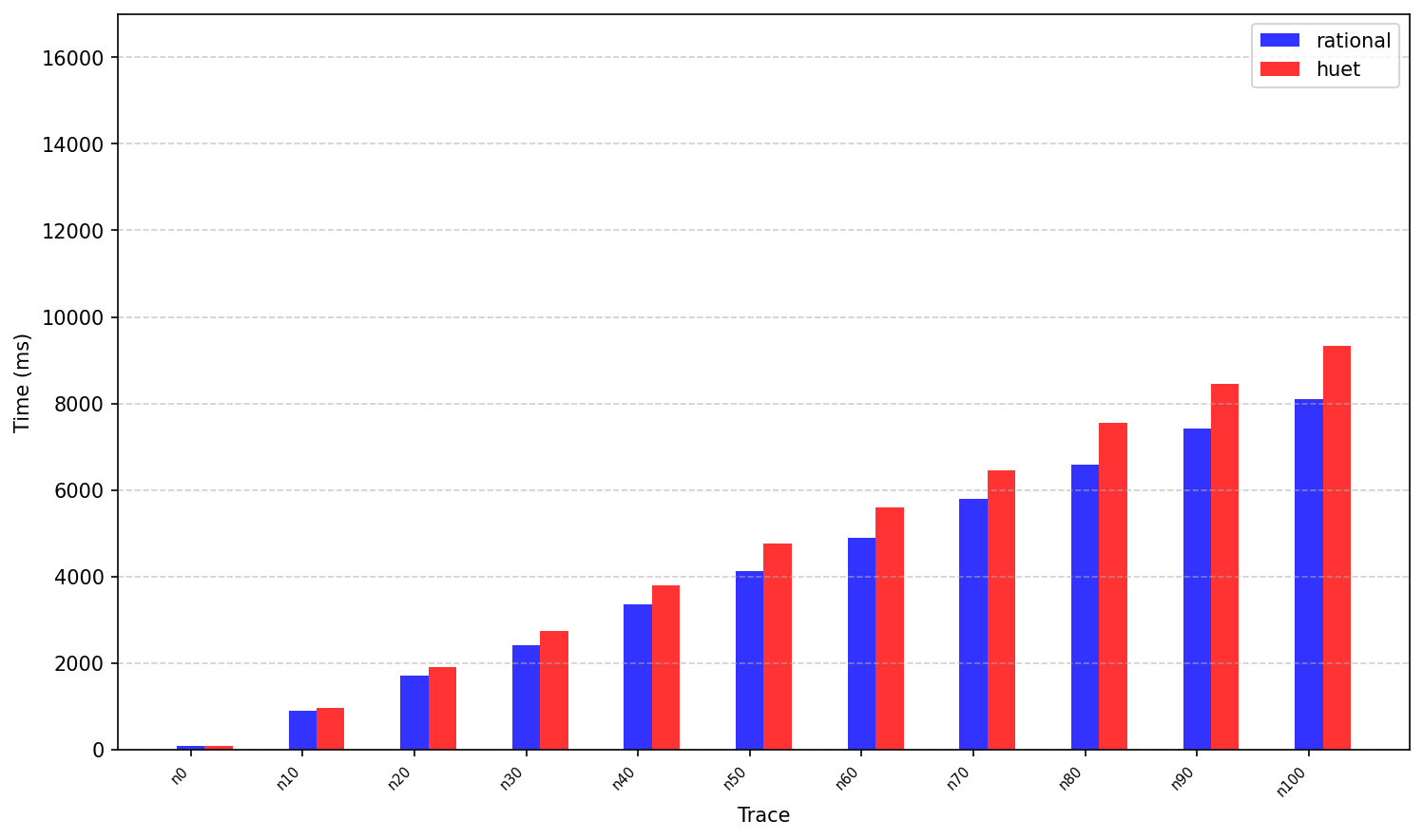}
        \subcaption{$\texttt{loops}_2$}
    \end{subfigure}
    \\
    \begin{subfigure}{.48\linewidth}
        \includegraphics[width=\linewidth]{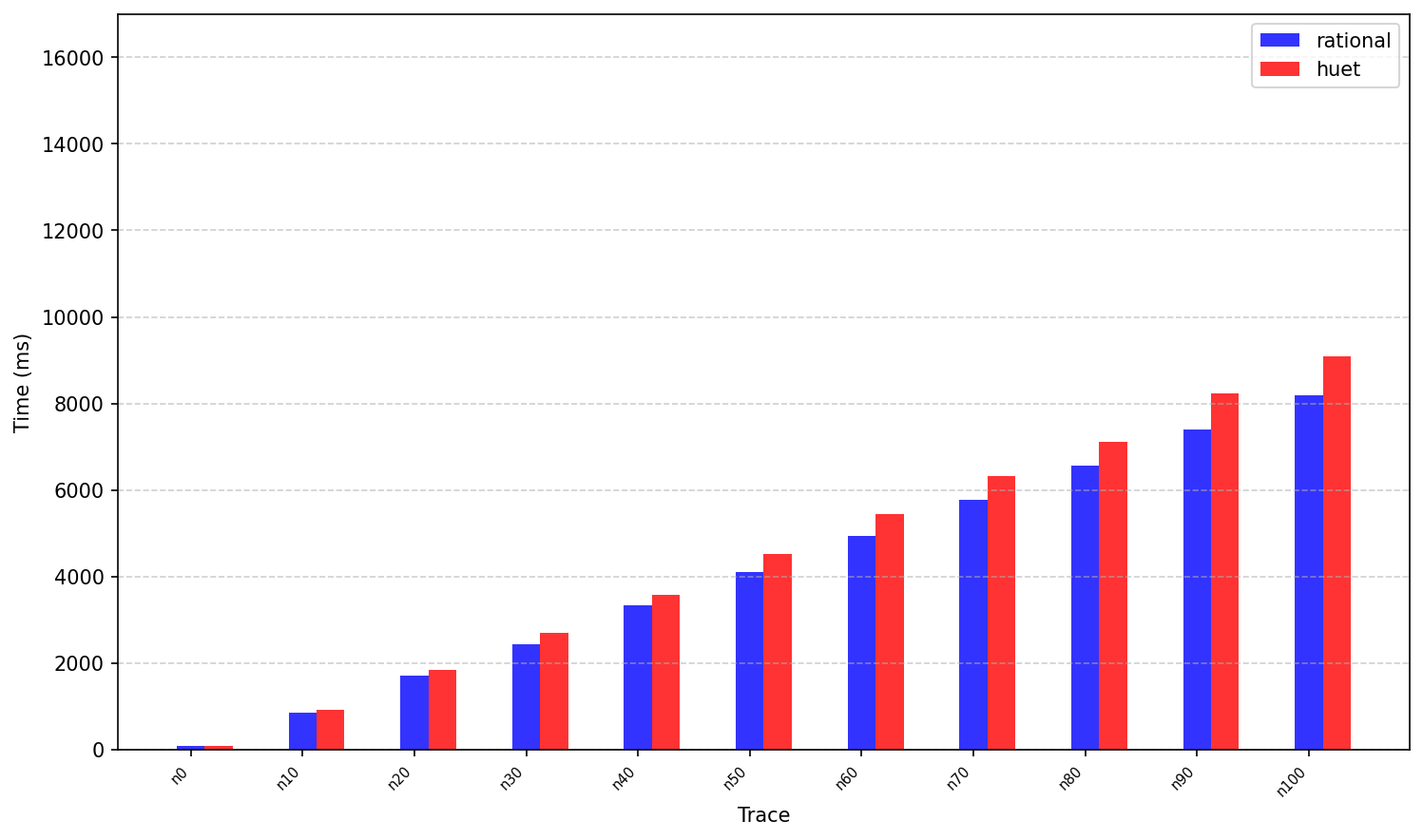}
        \subcaption{$\texttt{loops}_3$}
    \end{subfigure}
    \hfill
    \begin{subfigure}{.48\linewidth}
        \includegraphics[width=\linewidth]{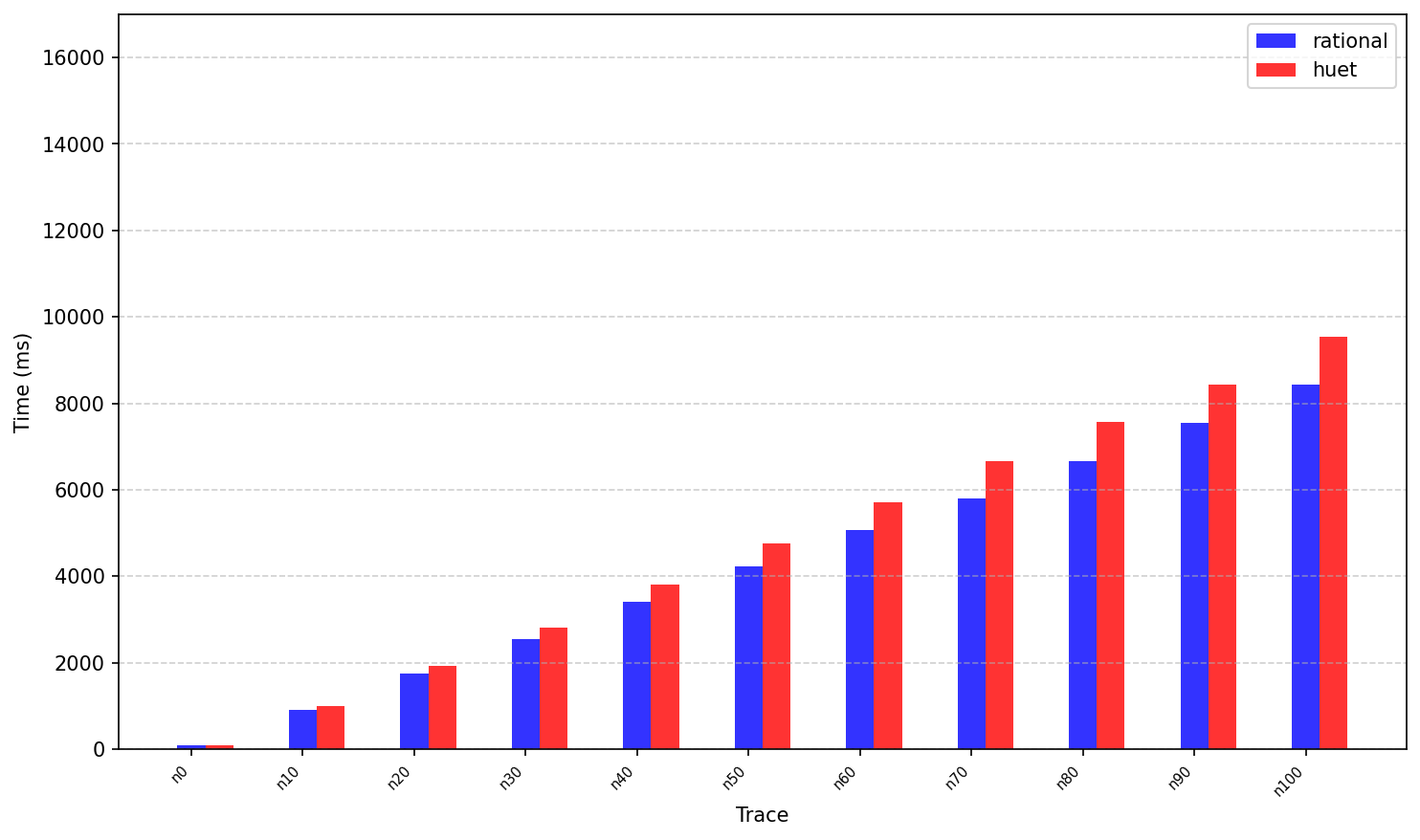}
        \subcaption{$\texttt{loops}_4$}
    \end{subfigure}
    \caption{Comparison of the ephemeral rational unification algorithm implementations on the synthetic tests}
    \label{fig:synth-ephemeral}
\end{figure}

\begin{figure}
    \centering
    \begin{subfigure}{.48\linewidth}
        \includegraphics[width=\linewidth]{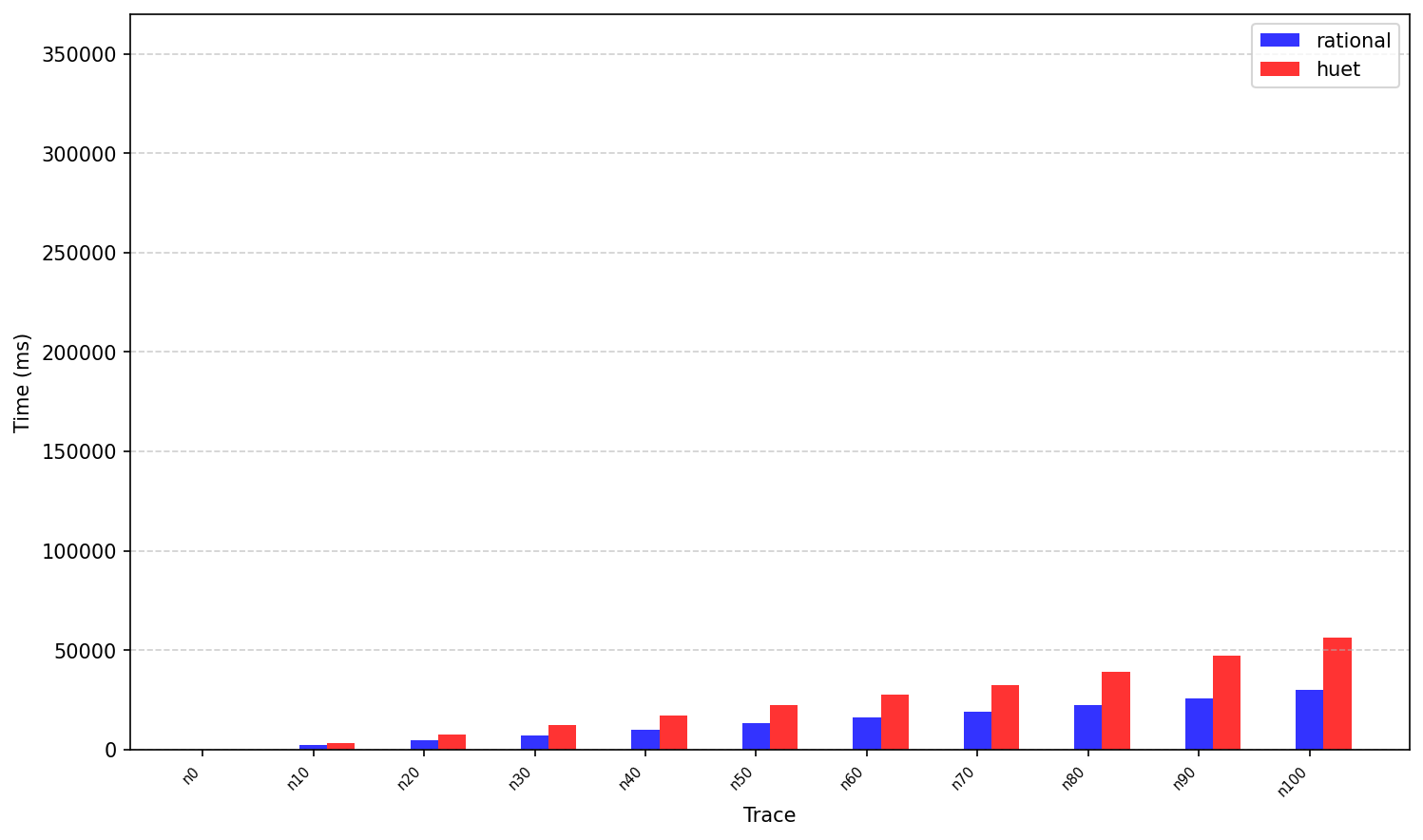}
        \subcaption{\texttt{full}}
    \end{subfigure}
    \hfill
    \begin{subfigure}{.48\linewidth}
        \includegraphics[width=\linewidth]{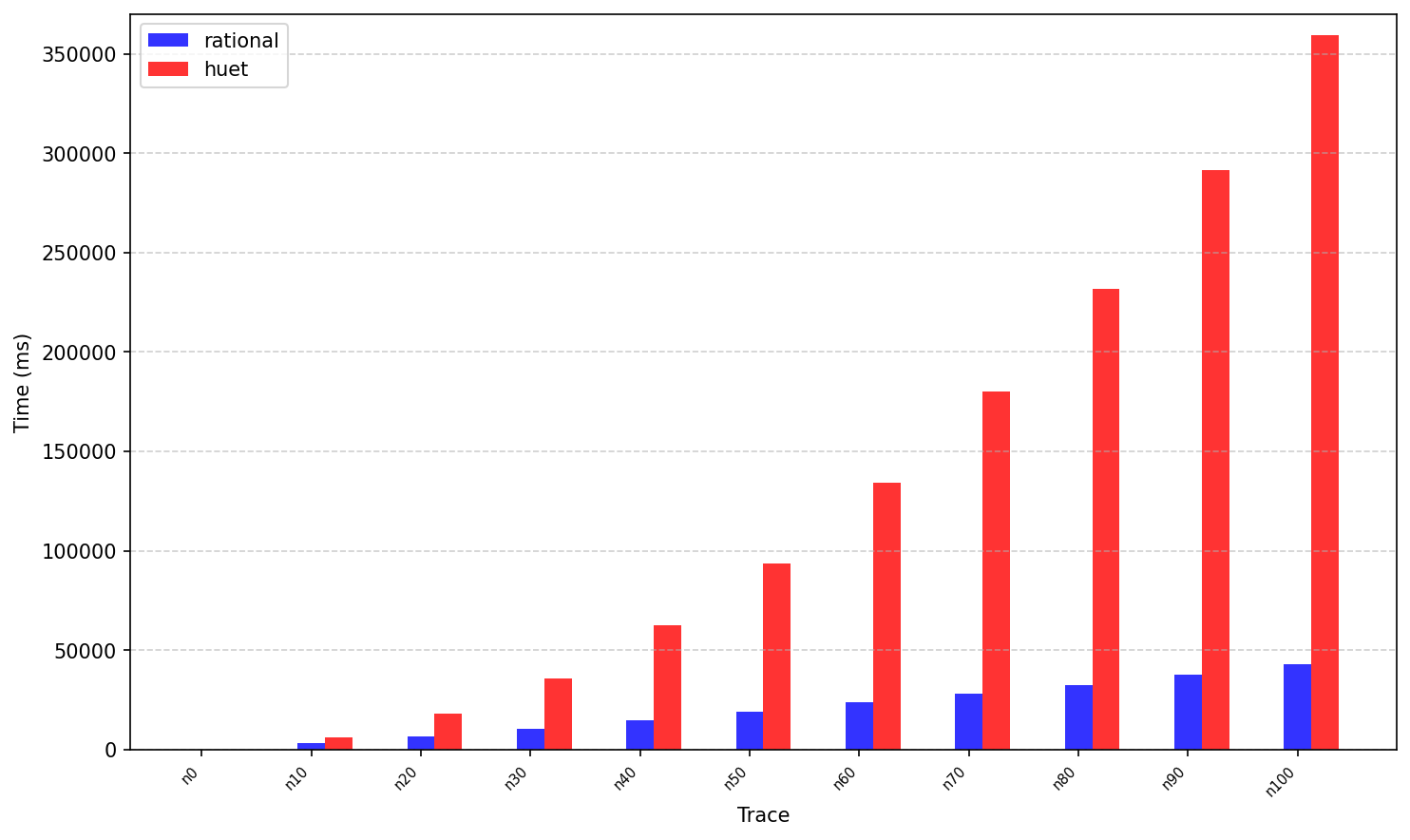}
        \subcaption{\texttt{left\_right}}
    \end{subfigure}
    \\
    \begin{subfigure}{.48\linewidth}
        \includegraphics[width=\linewidth]{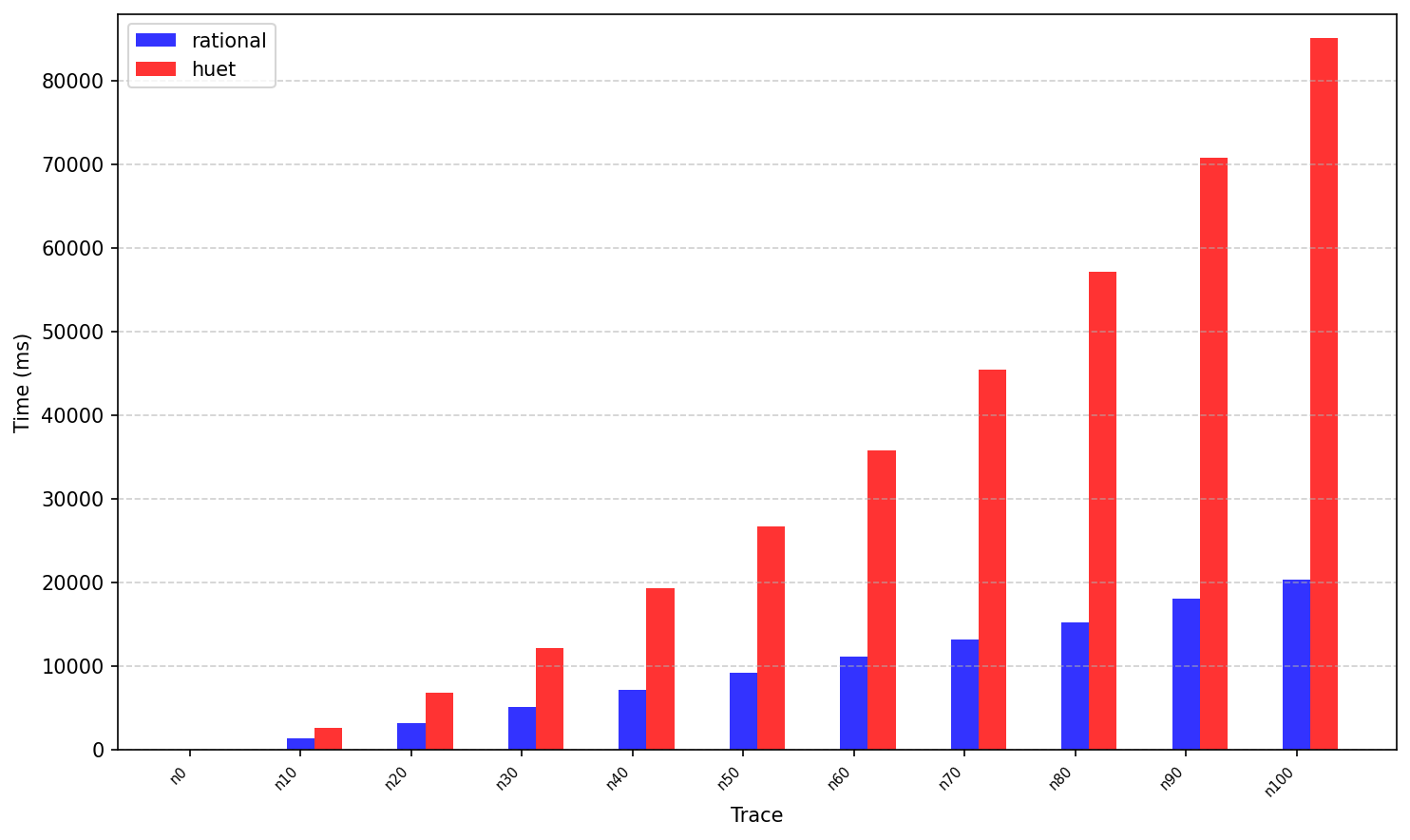}
        \subcaption{$\texttt{loops}_1$}
    \end{subfigure}
    \hfill
    \begin{subfigure}{.48\linewidth}
        \includegraphics[width=\linewidth]{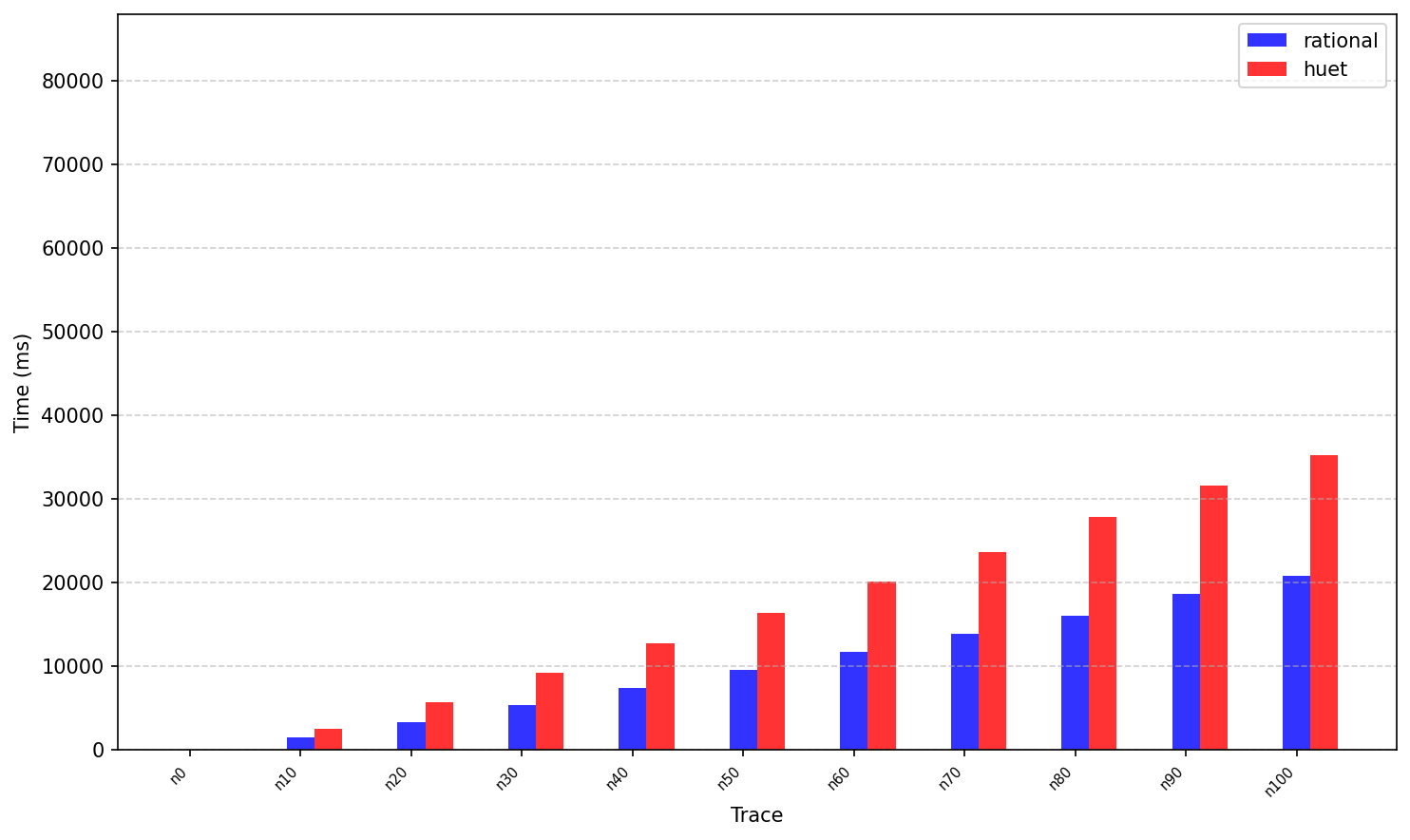}
        \subcaption{$\texttt{loops}_2$}
    \end{subfigure}
    \\
    \begin{subfigure}{.48\linewidth}
        \includegraphics[width=\linewidth]{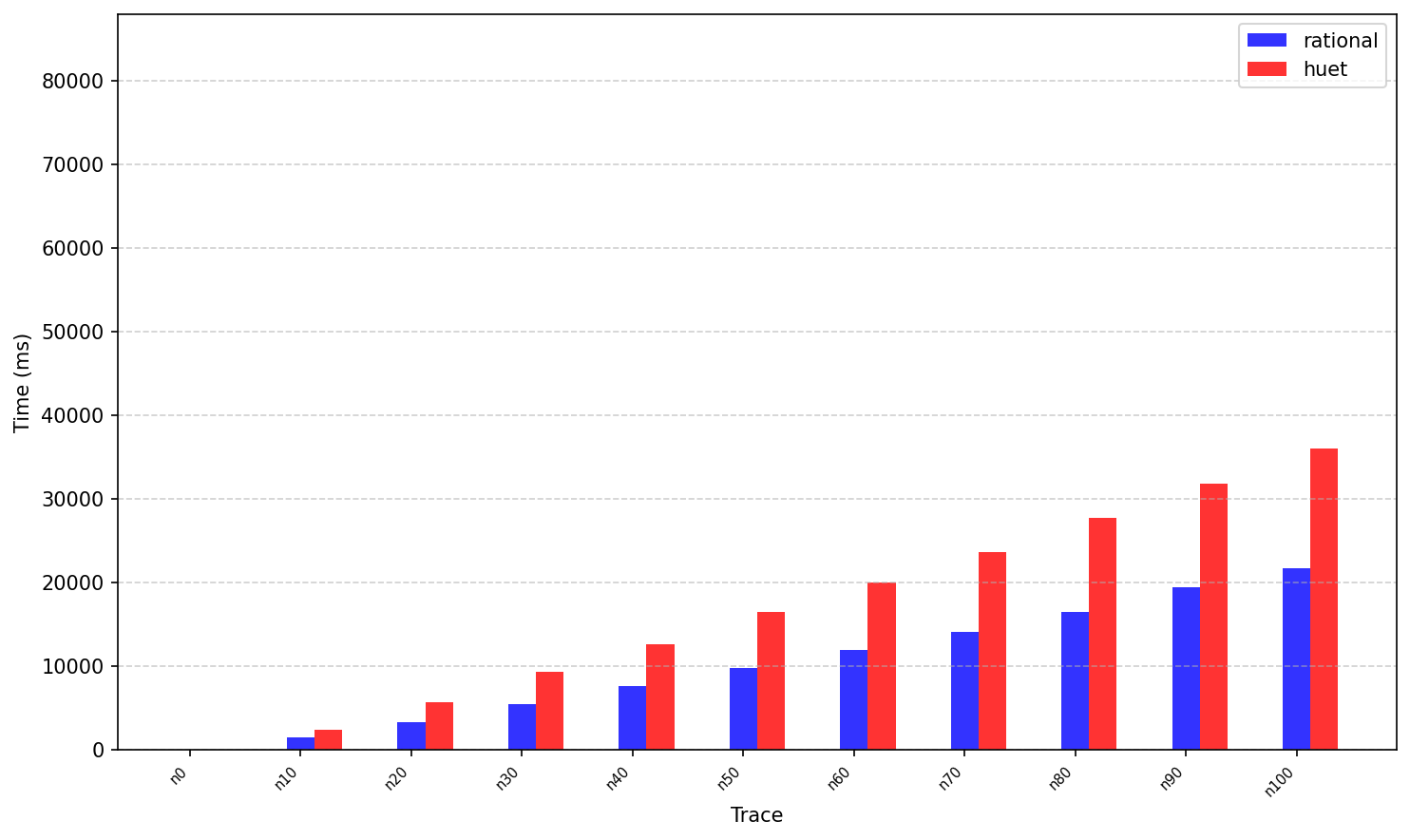}
        \subcaption{$\texttt{loops}_3$}
    \end{subfigure}
    \hfill
    \begin{subfigure}{.48\linewidth}
        \includegraphics[width=\linewidth]{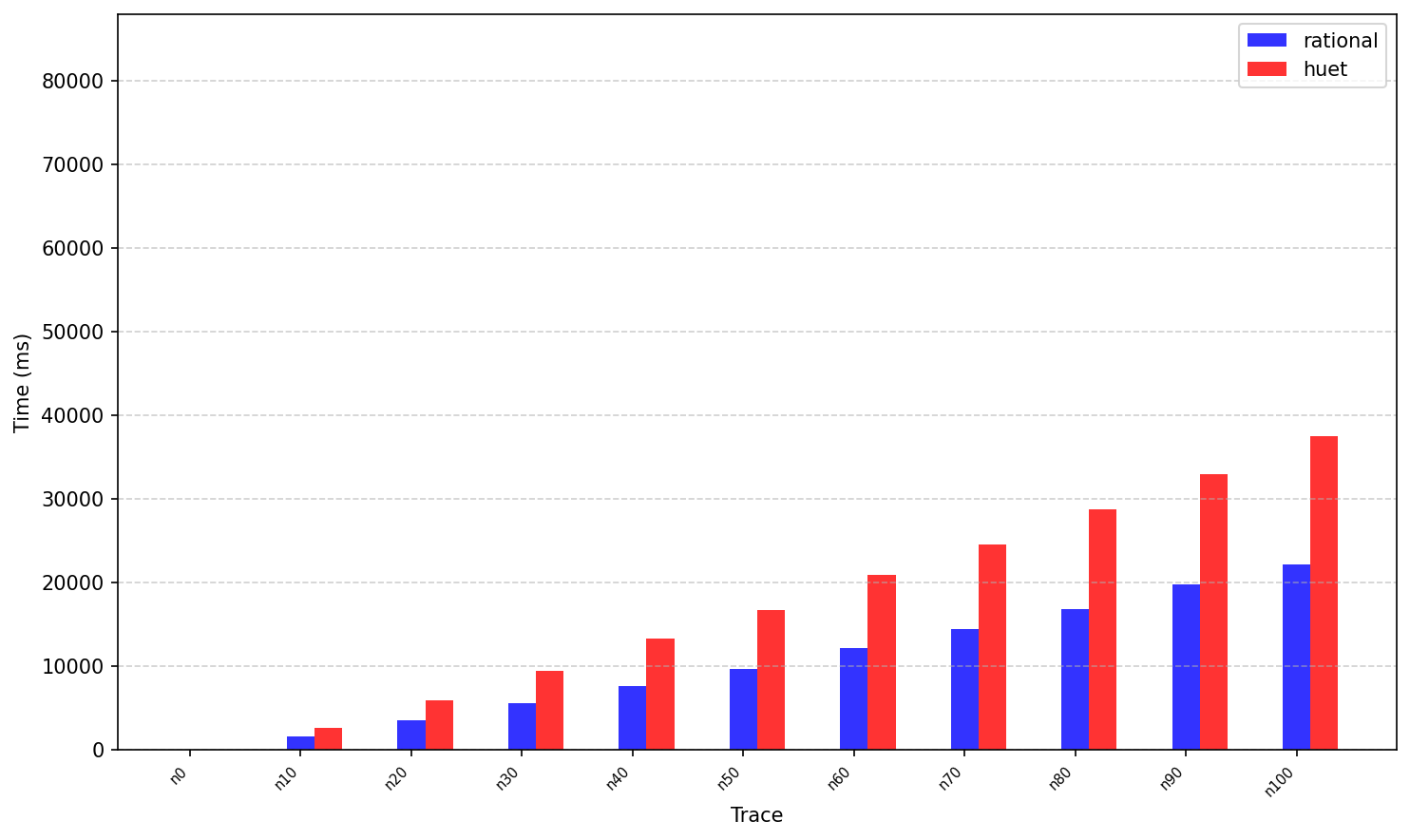}
        \subcaption{$\texttt{loops}_4$}
    \end{subfigure}
    \caption{Comparison of the persistent rational unification algorithm implementations on the synthetic tests}
    \label{fig:synth-persistent}
\end{figure}

The main evaluation results are:

\begin{itemize}
    \item Our algorithm performs \textbf{as fast as the conventional one} in the real-world scenarios. This is shown by \autoref{fig:baseline}.
    These plots show the algorithm's run time w.r.t. the conventional algorithm run time.
    As these plots show, our algorithm is always near to the main diagonal.
  
    \item While the Huet's algorithm works faster in ephemeral variant, it degrades being implemented in persistent settings (this is shown in \autoref{fig:baseline}, too, and in \autoref{fig:Lama}).
    \autoref{fig:synth-ephemeral} and \autoref{fig:synth-persistent} also show that even in the ephemeral implementation it works notably slower than our and moreover its performance depends on the ordering of equation sides.
\end{itemize}

The main (and perhaps the only) reason for the slowdown of Huet's algorithm is its approach to avoiding infinite loops. In theory, this algorithm assumes that every \textit{non-variable} term instance is attributed with a unique identifier. Under this assumption, it can accumulate equivalence classes for these term instances as for variables.

However, in practice, this is implemented by adding an extra slot that stores references between term instances. Since they are non-comparable, there is no way to perform an optimization like the variable-age heuristic which leads to both asymmetric running times (which is shown in the \autoref{fig:synth-ephemeral} and \autoref{fig:synth-persistent}) and longer ``dereference'' paths.

This also enlarges the object space in which equivalence information is accumulated since it must now contain both variables and term instances rather than just variables. While this is not a problem in ephemeral settings, in persistent ones it leads to a significant growth of intermediate states because we must preserve in them both equivalences between variables and between term instances.

In persistent settings additional time is required to allocate numeric identifiers for each non-variable term instance (because there is no other way to identify them in a persistent intermediate state). This is inconvenient in practical \miniKanren implementations and introduces an additional runtime overhead.

We also applied the Kruskal-Wallis test~\cite{kruskal1952use} and the Dunn post-hoc test~\cite{dunn1964multiple} with the Bonferroni correction~\cite{bonferroni1936teoria} to determine which algorithm works faster on which traces.
The overall summary is presented as numbers of statistically significant wins in \autoref{tab:analyse}. The best results are highlighted using the bold font.
The table shows again that the Huet's algorithm works better in the ephemeral implementation but loses its performance in the persistent one while our stays aligned with the conventional.

\begin{table}[b]
    \centering
    \begin{tabular}{l||c|c|c}
        \toprule
        \textbf{Algorithm} & \textbf{Ephemeral} & \textbf{Ephemeral with path compression} & \textbf{Persistent} \\
        \midrule
        Conventional & 755 & 915 & \textbf{1982} \\
        Our & 847 & 727 & \textbf{2155} \\
        Huet's & \textbf{2677} & \textbf{2651} & 62 \\
        \bottomrule
    \end{tabular}
    \caption{Statistically significant wins of every algorithm on all traces compared per-algorithm}
    \label{tab:analyse}
\end{table}

Also, the Huet's algorithm requires a more complex data structure representation than the conventional one while our doesn't.
These results show that our algorithm can be used as a drop-in replacement for the conventional one in \miniKanren implementations.

Additionally, following~\cite{domoratskiy2025empirical}, we have studied that path compression technique doesn't improve overall unification performance. This is shown in \autoref{tab:analyse-ephemeral}.

\begin{table}[b]
    \centering
    \begin{tabular}{l||c|c|c}
        \toprule
        \textbf{Variant} & \textbf{Conventional} & \textbf{Our} & \textbf{Huet's} \\
        \midrule
        Ephemeral & \textbf{597} & \textbf{784} & 494 \\
        Ephemeral with path compression & 470 & 345 & \textbf{685} \\
        \midrule
        Median ratio (avg) & 1.02 & 1.00 & 1.02 \\
        \bottomrule
    \end{tabular}
    \caption{Statistically significant wins of every algorithm on all traces comparing the ephemeral variants}
    \label{tab:analyse-ephemeral}
\end{table}

\section{Related Work}

Unification of rational terms is a well-studied area in computational logic. Although there are many
works~\cite{huet1976resolution, martelli1984efficient, mukai1983unification, jaffar1984efficient} presenting a
first-order unification algorithms for rational terms they all follow one of two possible approaches: either as
in the Huet's algorithm~\cite{huet1976resolution} or as in Martelli and Rossi's algorithm~\cite{martelli1984efficient}.
The first approach is to accumulate a set of equivalences between input terms and the second is to accumulate only a
set of equivalences between input variables.

We observe that only Huet-like approach is used\footnote{\url{https://github.com/SWI-Prolog/swipl/blob/45e83178379f0fd6f918bd294bb361f3db7bf1d7/src/pl-prims.c\#L71}}~\cite{gauthier2004numbering} in practice since it proved itself efficient in ephemeral setting. At the same time, we haven't seen yet
any practical applications of the second approach which we have shown to be preferable for persistent rational unification.

Also, although all works on first-order rational unification discuss time complexity, none of them present evaluation of the algorithms on real data. In our work
we provide the comprehensive and verifiable comparative evaluation for two known approaches, and compare them with the most used conventional algorithm.

The our previous work~\cite{domoratskiy2025empirical} provide an inefficient algorithm which was roughly derived from the Martelli and Rossi's approach and moreover broke the MGU property in general. The latter was addressed with the introduction of the \commonpart operation rather than fresh variable generation for each non-variable term.

\section{Conclusion}

We presented a rational unification algorithm which could be used as a drop-in replacement for the conventional one to support recursive data structures in \textsc{miniKanren}.
We have proven its correctness and termination properties in \Rocq and evaluated it on a real-world data and synthetic tests.

The future work may include more research on the usage of the algorithm to derive non-rational unification without explicit ``occurs check''. It is also important to research
applications of ``inductive''~\cite{bol1991analysis} and ``coinductive''~\cite{simon2006coinductive} guards for relations which are required to improve the expressivity of
relational programs dealing with rational terms.

\bibliographystyle{plain}
\bibliography{main}

\end{document}